\RequirePackage[mathlines]{lineno}
\documentclass[amsmath,amssymb,twocolumn,prd,floatfix,showpacs, nofootinbib]{revtex4-1}
\usepackage{tabularx}  % tabular environment
\usepackage{bm, color} % math bold and /color
\usepackage{overpic,subfigure} % figures
\usepackage{multirow}
\usepackage{array}
\usepackage{dcolumn} % to deal with numbers with units and uncertainties
\usepackage[symbol]{footmisc}
\usepackage{epstopdf}
\usepackage{lineno}
\usepackage{amsmath}
\usepackage{epstopdf}
\usepackage{hyperref}

\newcommand{\BESIIIorcid}[1]{\href{https://orcid.org/#1}{\hspace*{0.1em}\raisebox{-0.45ex}{\includegraphics[width=1em]{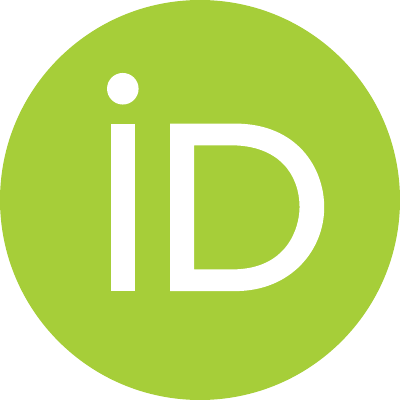}}}}

\hypersetup{hypertex=true,
	        colorlinks = true,
            linkcolor = blue,
            urlcolor = blue,
            citecolor = blue}
\begin{document}
%\linenumbers

\title{\boldmath Measurements of the branching fractions of $\chi_{cJ}\to \phi\phi\eta, \phi\phi\eta^{\prime}$ and $\phi K^+K^-\eta$ }

\author{
%% Saved at => 2025-07-28
M.~Ablikim$^{1}$\BESIIIorcid{0000-0002-3935-619X},
M.~N.~Achasov$^{4,c}$\BESIIIorcid{0000-0002-9400-8622},
P.~Adlarson$^{82}$\BESIIIorcid{0000-0001-6280-3851},
X.~C.~Ai$^{87}$\BESIIIorcid{0000-0003-3856-2415},
R.~Aliberti$^{39}$\BESIIIorcid{0000-0003-3500-4012},
A.~Amoroso$^{81A,81C}$\BESIIIorcid{0000-0002-3095-8610},
Q.~An$^{78,64,\dagger}$,
Y.~Bai$^{62}$\BESIIIorcid{0000-0001-6593-5665},
O.~Bakina$^{40}$\BESIIIorcid{0009-0005-0719-7461},
Y.~Ban$^{50,h}$\BESIIIorcid{0000-0002-1912-0374}
% ...
\\
\vspace{0.2cm}
(BESIII Collaboration)\\
\vspace{0.2cm} {\it
$^{1}$ Institute of High Energy Physics, Beijing 100049, People's Republic of China\\
% ...
$^{4}$ Budker Institute of Nuclear Physics SB RAS (BINP), Novosibirsk 630090, Russia\\
% ...
$^{39}$ Johannes Gutenberg University of Mainz, Johann-Joachim-Becher-Weg 45, D-55099 Mainz, Germany\\
$^{40}$ Joint Institute for Nuclear Research, 141980 Dubna, Moscow region, Russia\\
% ...
$^{50}$ Peking University, Beijing 100871, People's Republic of China\\
% ...
$^{62}$ Southeast University, Nanjing 211100, People's Republic of China\\
% ...
$^{64}$ State Key Laboratory of Particle Detection and Electronics, Beijing 100049, Hefei 230026, People's Republic of China\\
% ...
$^{78}$ University of Science and Technology of China, Hefei 230026, People's Republic of China\\
% ...
$^{81}$ University of Turin and INFN, (A)University of Turin, I-10125, Turin, Italy; (B)University of Eastern Piedmont, I-15121, Alessandria, Italy; (C)INFN, I-10125, Turin, Italy\\
$^{82}$ Uppsala University, Box 516, SE-75120 Uppsala, Sweden\\
% ...
$^{87}$ Zhengzhou University, Zhengzhou 450001, People's Republic of China\\
\vspace{0.2cm}
$^{\dagger}$ Deceased\\
% ...
$^{c}$ Also at the Novosibirsk State University, Novosibirsk, 630090, Russia\\
% ...
$^{h}$ Also at State Key Laboratory of Nuclear Physics and Technology, Peking University, Beijing 100871, People's Republic of China\\
% ...
}
%% ends here %%
}

\begin{abstract}
  Using a sample of $(2712.4 \pm 14.3)\times 10^6 ~\psi$(3686) events collected by the BESIII detector at the BEPCII collider, we measure the branching fractions of the decays $\chi_{cJ}\to \phi\phi\eta,~\phi\phi\eta^{\prime}$, and~$\phi K^+K^-\eta $ ($J = 0, 1, 2$). The obtained branching fractions are
$\mathcal{B}(\chi_{c0} \to \phi\phi\eta)         =   (7.40 \pm 0.23 \pm 0.55)\times10^{-4}$,
$\mathcal{B}(\chi_{c1} \to \phi\phi\eta)         =   (3.33 \pm 0.14 \pm 0.25)\times10^{-4}$,
$\mathcal{B}(\chi_{c2} \to \phi\phi\eta)         =   (5.46 \pm 0.17 \pm 0.40)\times10^{-4}$,
$\mathcal{B}(\chi_{c0} \to \phi\phi\eta^\prime)  =   (2.96 \pm 0.23 \pm 0.29)\times10^{-4}$,
$\mathcal{B}(\chi_{c1} \to \phi\phi\eta^\prime)  =   (0.69 \pm 0.10 \pm 0.08)\times10^{-4}$,
$\mathcal{B}(\chi_{c2} \to \phi\phi\eta^\prime)  =   (0.65 \pm 0.09 \pm 0.07)\times10^{-4}$,
$\mathcal{B}(\chi_{c0} \to \phi K^+K^-\eta)      =   (1.23 \pm 0.08 \pm 0.10)\times10^{-4}$,
$\mathcal{B}(\chi_{c1} \to \phi K^+K^-\eta)      =   (1.00 \pm 0.07 \pm 0.07)\times10^{-4}$, and
$\mathcal{B}(\chi_{c2} \to \phi K^+K^-\eta)      =   (1.82 \pm 0.09 \pm 0.14)\times10^{-4}$,
where $K^+K^-$ is not from the decay of a $\phi$ meson, the first uncertainties are statistical and the second systematic.
The branching fractions of $\chi_{cJ}\to \phi\phi\eta$ are measured with precision improved by factors of $1.5-1.9$, and those of $\chi_{cJ}\to \phi\phi\eta^\prime$ and $\phi K^+K^-\eta$ are measured for the first time.
\end{abstract}

\maketitle
%\linenumbers

\section{Introduction}

Studies of the properties of $c \bar{c}$ states play an important role in understanding the interplay between perturbative and non-perturbative effects in quantum chromodynamics (QCD).
Below the open charm threshold, both $J/\psi$ and $\psi$(3686) mesons mainly decay into light hadrons through the annihilation of the $c \bar{c}$ pair into three gluons or one single virtual photon, with the decay width proportional to the modulus squared of the charmonium wave function~\cite{wave-fucntion}. QCD has been tested thoroughly at high energy region where the strong interaction coupling constant is small. However, in the low energy region, theoretical calculations based on first principles of QCD are still unreliable since the non-perturbative contribution is significant, and various effective field theories are introduced~\cite{field-1, field-2, field-3} to approximate these non-perturbative contributions.
%-------
In addition to the decays of $J/\psi$ and $\psi$(3686)~\cite{ref::pdg2024}, the decays of the $\chi_{cJ}$ states (with $J$ = 0, 1, 2)~\cite{chicj-1,chicj-2} are also valuable for probing a wide range of QCD phenomena.

To date, only a few decays of $\chi_{cJ} \to VVP$~\cite{ref::pdg2024} ($V$ and $P$ represent vector and pseudoscalar mesons, respectively) have been studied, and no branching fraction measurement for $\chi_{cJ} \to \phi\phi\eta^{\prime}$ has been reported. %------------------------------------------------
Experimental study of the decays of $\chi_{cJ}$ into these final states is crucial for understanding the internal charmonium structure and testing the phenomenological mechanisms of non-perturbative QCD.

With $\chi_{cJ}$ mesons abundantly produced in the radiative decays of $\psi$(3686)~\cite{ref::pdg2024}, the BESIII experiment provides an ideal environment for searching for new $\chi_{cJ}$ decays. By analyzing $(2712.4 \pm 14.3)\times 10^6 ~\psi$(3686) events~\cite{psip-number} collected by the BESIII detector, we report the measurements of the branching fractions for $\chi_{cJ}\to \phi\phi\eta$ with much improved precision~\cite{BESIII:2019ngm}, as well as the first measurements of the branching fractions for $\chi_{cJ}\to \phi\phi\eta^\prime$ and $\chi_{cJ}\to \phi K^+K^-\eta$.
Throughout this paper,  the $K^+K^-$ from the decay of $\phi$ mesons is excluded in the signal $\chi_{cJ}\to \phi K^+K^-\eta$.
%Throughout this paper, in any signal decay the $K^+K^-$ is not from the decay of $\phi$ mesons.

\section{BESIII DETECTOR AND MONTE CARLO SIMULATION}
\label{sec:BES}
The BESIII detector~\cite{ref::detector} records symmetric $e^+e^-$ collisions provided by the BEPCII storage ring~\cite{ref::collider}
in the center-of-mass energy $(\sqrt{s})$ range from 1.84 to 4.95~GeV, with a peak luminosity of $1.1 \times 10^{33}\;\text{cm}^{-2}\text{s}^{-1}$ achieved at $\sqrt{s} = 3.773\;\text{GeV}$.
BESIII has collected large data samples in this energy region~\cite{Ablikim:2019hff}.

The cylindrical core of the BESIII detector covers 93\% of the full solid angle and consists of a helium-based multilayer drift chamber~(MDC),
a plastic scintillator time-of-flight system~(TOF), and a CsI~(Tl) electromagnetic calorimeter~(EMC), which are all enclosed in a superconducting
solenoidal magnet providing a 1.0~T magnetic field. The solenoid is supported by an octagonal flux-return yoke with resistive plate counter muon
identification modules interleaved with steel. The charged-particle momentum resolution at $1~{\rm GeV}/c$ is $0.5\%$, and the ${\rm d}E/{\rm d}x$
resolution is $6\%$ for electrons from Bhabha scattering. The EMC measures photon energies with a resolution of $2.5\%$ ($5\%$) at $1$~GeV in the barrel
(end-cap) region. The time resolution in the TOF barrel region is 68~ps, while that in the end-cap region was 110~ps.
The end-cap TOF system was upgraded in 2015 using multi-gap resistive plate chamber technology, providing a time resolution of 60~ps~\cite{Tof1,Tof2,Tof3}.
Around $83\%$ of the data used  in this analysis benefits from this upgrade.

Simulated data samples produced with a {\sc geant4}-based~\cite{Geant4} Monte Carlo (MC) package, which includes the geometric description of the BESIII detector and the detector response, are used to determine detection efficiencies and to estimate background. The simulation models the beam energy spread and initial state radiation (ISR) in the $e^+e^-$ annihilations with the generator {\sc kkmc}~\cite{Jadach01}.
%----
The inclusive MC sample includes the production of the $\psi(3686)$ resonance, the ISR production of the $J/\psi$, and the continuum processes incorporated in {\sc kkmc}~\cite{Jadach01}.
%------
All particle decays are modelled with {\sc evtgen}~\cite{Lange01,Lange02} using branching fractions either taken from the
Particle Data Group (PDG)~\cite{ref::pdg2024}, when available, or otherwise estimated with {\sc lundcharm}~\cite{Lundcharm00}. Final state radiation from charged final state particles is incorporated using the {\sc photos} package~\cite{PHOTOS}.
An inclusive MC sample containing $2.7\times10^{9}$ generic $\psi(3686)$ events is used to study background.

\section{EVENT SELECTION}
\label{sec:selection}

The charged tracks detected in the MDC are required to be within a polar angle ($\theta$) range of $|\rm{cos\theta}|<0.93$,  where $\theta$ is defined with respect to the $z$-axis,
which is the symmetry axis of the MDC.
The distance of closest approach to the interaction point (IP)  must be less than 10\,cm along the $z$-axis
and less than 1\,cm in the transverse plane.
%--------------
Particle identification~(PID) for charged tracks combines measurements of the energy deposited in the MDC and the flight time in the TOF to form likelihoods $\mathcal{L}(h)$~($h$ = $\pi$ and $K$) for each hadron $h$ hypothesis. The charged tracks with $\mathcal{L}(K)>\mathcal{L}(\pi)$ are assigned as kaons, while those with $\mathcal{L}(\pi)>\mathcal{L}(K)$ is assigned as pions.

Photon candidates are identified using showers in the EMC. The deposited energy of each shower must be more than 25~MeV in the barrel region ($|\cos \theta|< 0.80$),
and more than 50~MeV in the end-cap region ($0.86 <|\cos \theta|< 0.92$).
To exclude showers that originate from interactions between charged tracks and crystals, the angle subtended by the EMC shower and the position of the closest charged track at the EMC must be greater than 10 degrees as measured from the IP. To suppress electronic noise and showers unrelated to the event, the difference between the EMC time and the event start time is required to be within [0, 700]\,ns.

The signal events are reconstructed from the charmonium transitions
$\psi(3686) \to \gamma\chi_{cJ}$, $\chi_{cJ} \to \phi\phi\eta,~\phi\phi\eta^{\prime}$, and $\phi K^+K^-\eta $, where
%-------
the $\phi$ meson is reconstructed via $\phi \to K^+K^-$ and
%--------------
the $\eta$ meson is reconstructed via $\eta \to \gamma\gamma$.
The $\gamma\gamma$ pair with the invariant mass $M_{\gamma\gamma}$ closest to the $\eta$ nominal mass~\cite{ref::pdg2024} is selected if there are multiple combinations. The $\eta$ signal region is defined as $0.51\leq M_{\gamma\gamma} \leq0.57$~GeV/$c^{2}$.
%----------
The $\eta^{\prime}$ meson is reconstructed by two decay modes, $\eta^{\prime} \to \pi^+\pi^-\gamma$ (Mode I) and $\eta^{\prime} \to \pi^+\pi^-\eta$ with $\eta \to \gamma\gamma$ (Mode II). Candidate events with $|M_{\pi^+\pi^-\gamma~\rm{or}~\pi^+\pi^-\eta}-0.958|<0.015$~GeV/$c^{2}$ is defined as the $\eta^{\prime}$ signal region.
%-------------.
For $\chi_{cJ} \to \phi\phi\eta$ and $\phi K^+K^-\eta $, candidate events are required to have four charged tracks with zero net charge and at least three photons.
For $\chi_{cJ} \to \phi\phi\eta^{\prime}$, candidate events are required to have six charged tracks with zero net charge and at least two (three) photons for Mode I (II).

A four constraints (4C) kinematic fit is applied to the events under the hypothesis of $\psi(3686) \to \gamma\chi_{cJ}, \chi_{cJ} \to$ signal, by imposing four-momentum conservation.
In each event, if more than one combination survives the selections, the one with the smallest  $\chi_{\rm 4C}^{2}$ of the kinematic fit is retained. Meanwhile, the candidate events are required to satisfy $\chi^2_{\rm 4C}<38$. It is optimized with the Figure-of-Merit $\mathit{S}/\sqrt{\mathit{S}+\mathit{B}}$, where $S$ is the number of signal events from the signal MC samples, $B$ is the number of background events from the inclusive MC samples, and both of them have been scaled to the data size.
%For $\chi_{cJ} \to \phi\phi\eta^{\prime}$, we continue to use the $\chi^2_{\rm 4C}$ value above due to the small statistic.

\begin{figure}[htbp]
	\centering
    \includegraphics[width=1.0\linewidth]{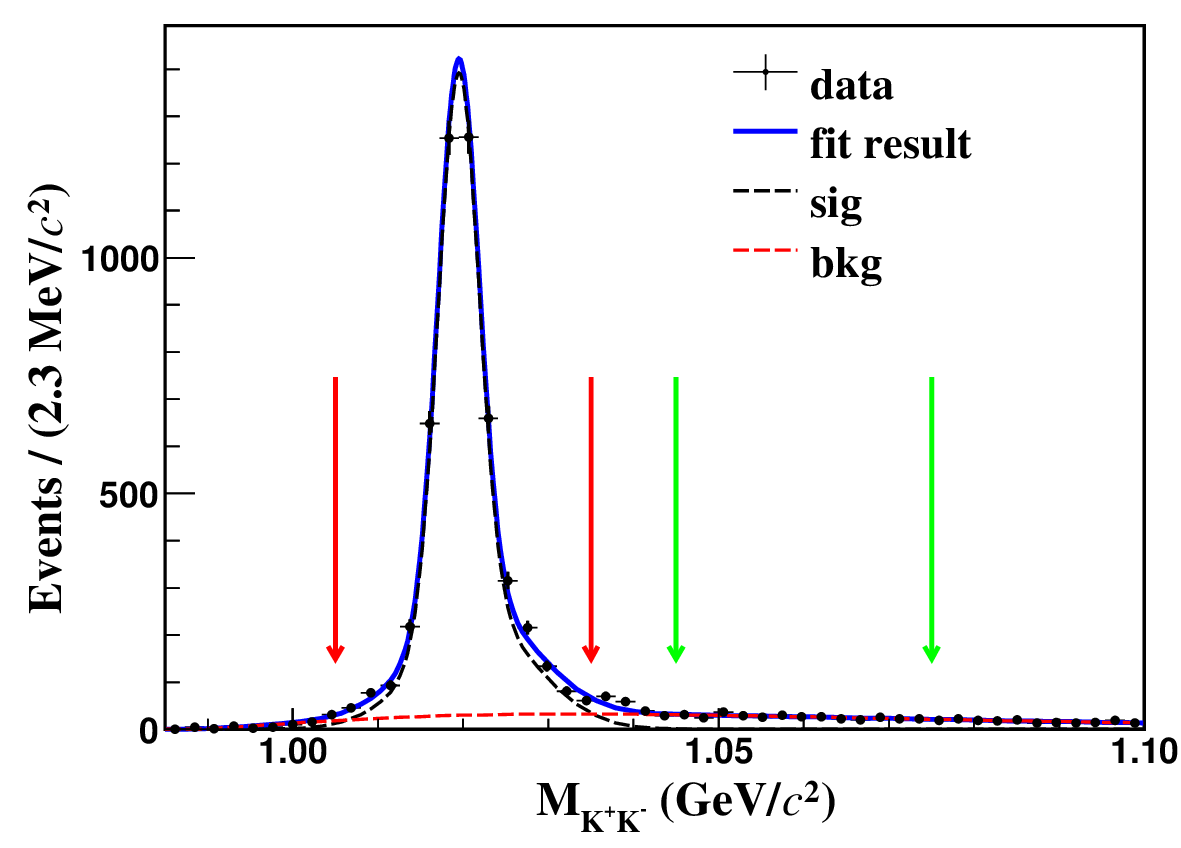}
	\caption {The $M_{K^+K^-}$ distribution. The dots with error bars are data, the blue solid curve is the fit result, the red dashed curve is the fitted background, and the black dashed curves are the fitted signal shapes. The red arrows show the $\phi$ signal region and the green ones show the $\phi$ sideband region. The events are from the accepted candidates of $\chi_{cJ} \to \phi\phi\eta$.}
	\label{fig:dataphi-1}
\end{figure}

\begin{figure}[htbp]
	\centering
	\includegraphics[width=1.0\linewidth]{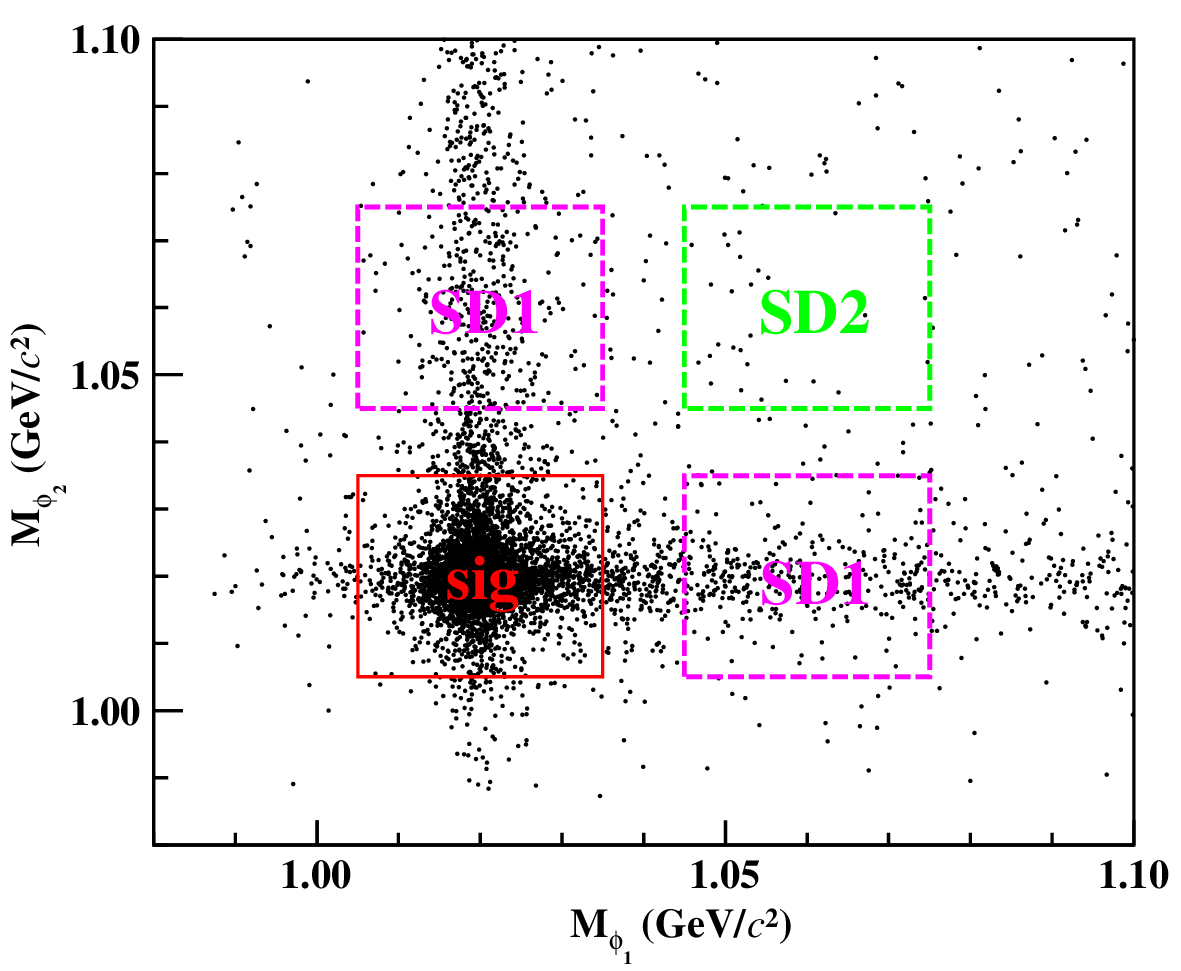}
	\caption {The distribution of $M_{\phi_2}$ versus $M_{\phi_1}$ of the accepted candidates of $\chi_{cJ} \to K^+K^+K^-K^-\eta$, with the $\eta$ candidates lying in the $\eta$ signal region in data. The red box shows the $\phi\phi$ signal region, the pink boxes show the SD1 regions, and the green one shows the SD2 region.}
	\label{fig:phi2D-1}
\end{figure}

For the decays $\chi_{cJ} \to \phi\phi\eta$ and $\phi\phi\eta^{\prime}$, the $\phi\phi$ combinations are chosen by requiring
% from the combination with
the minimum value of $\Delta M^2=(M_{K_i^+K_j^-}-m_\phi)^2+(M_{K_{1-i}^+K_{1-j}^-}-m_\phi)^2$, where $M_{K^+K^-}$ is the invariant mass of the $K^+K^-$ combination, $m_{\phi}$ is the $\phi$ nominal mass~\cite{ref::pdg2024}, and $i, j$ can be 0 or 1. In each event, one of the $\phi$ candidates is randomly chosen to be $\phi_{1}$ with the other $\phi_{2}$.
%MC studies show that the miscombination rates for both $\eta$ and $\phi$ candidates are no more than 1.66$\%$.
 The $\phi$ signal region is set to be within $3\sigma$, $1.005\leq M_{K^+K^-} \leq1.035$~GeV/$c^{2}$. Figure~\ref{fig:dataphi-1} shows the result of the fit to the $M_{K^+K^-}$ distribution obtained when one of the two combinations is randomly selected. In the fit, the signal shape is described by a double Gaussian function, and the background shape is represented by an ARGUS function~\cite{Argus}. The two-dimensional (2D) $\phi\phi$ signal region is shown as the ``sig" box in Fig.~\ref{fig:phi2D-1}, where $M_{\phi_1}$ and $M_{\phi_2}$ denote the invariant masses of the randomly assigned $\phi_1$ and $\phi_2$, respectively.
 Two kinds of 2D $\phi\phi$ sideband regions, named ``SD1" and ``SD2", correspond to events with one correct $\phi$ candidate and one wrong $\phi$ candidate and events with two wrong $\phi$ candidates, respectively.
 %----------------------------
 For the decays $\chi_{cJ} \to \phi K^+K^-\eta$, all of the $K^+K^-$ combinations are kept. The signal region is defined as $M_{\phi_{1~(2)}} \in (1.005, 1.035)$~GeV/$c^{2}$,$M_{\phi_{2~(1)}}> 1.09$~GeV/$c^{2}$. The sideband region is defined as $M_{\phi_{1~(2)}} \in (1.045, 1.075)$~GeV/$c^{2}$, $M_{\phi_{2~(1)}}> 1.09$~GeV/$c^{2}$.
 %------------------------------------------------
 %For the decays $\chi_{cJ} \to K^+K^+K^-K^-\eta$, we require all $M_{K^+K^-}$ combinations to be greater than 1.065~GeV/$c^{2}$.

\section{BACKGROUND AND SIGNAL YIELDS}
The potential background components from $\psi$(3686) decays are studied by analyzing the inclusive MC sample with the generic event type analysis tool, TopoAna~\cite{topo}.
No significant peaking background is observed.
%----------------------------------------------------------
For the decays $\chi_{cJ} \to \phi K^+K^-\eta$,  we require $|M_{K^+K^+K^-K^-}-M_{J/\psi}|>0.015$~GeV/$c^{2}$ in order to suppress the background from $\psi(3686) \to \eta J/\psi, J/\psi \to K^+K^+K^-K^-$.
%----------------------------------------------------------
In order to suppress the background from $\psi(3686) \to K^+K^+K^-K^-\pi^+\pi^-\pi^0$ in $\chi_{cJ} \to \phi\phi\eta^{\prime}_{\pi^+\pi^-\gamma}$ candidate events, $|M_{\gamma\gamma}-M_{\pi^0}|>0.015$~GeV/$c^{2}$ is applied.
%-----------------------------------------------------------
Furthermore, the possible continuum background contribution is investigated by examining the data sample taken at $\sqrt{s} = $ 3.650 GeV, corresponding to an integrated luminosity of 454.33 pb$^{-1}$~\cite{lum}. Since only a few events survive the selection criteria, the contribution from continuum events  is ignored.

To determine signal yields, simultaneous 2D unbinned maximum likelihood fits are performed to the distribution of $M_{\chi_{cJ}}$ versus $M_{\gamma\gamma}$ for the candidate events from the signal and sideband regions.
%----------
In the 2D fit, the signal shape of the $\eta$ peak is described by the MC-simulated shape,
%----
and the signal shapes of the $\chi_{cJ}$ peaks are described by MC simulation shape corrected by a phase space factor and a damping factor.
%----------------------------
The phase space factor is defined as (\( PS(\sqrt{s}) = (\frac{E_{\gamma}}{E^0_{\gamma}})^{3/2}\))~\cite{E1-1, E1-2}, where $E_{\gamma}$ is the radiative photon energy and $E^0_{\gamma}$ corresponds to the photon energy at the $\chi_{cJ}$ nominal mass.
The damping factor \(D(\sqrt{s}) = ( \frac{(E^0_{\gamma})^2}{E^0_{\gamma} E_{\gamma} + (E^0_{\gamma} - E_{\gamma})^2} )^{1/2} \)~\cite{dam-1} is introduced to suppress the higher energy tail.
%--------------------------------------
The background shapes for both $M_{\chi_{cJ}}$ and $M_{\gamma\gamma}$ are described by Chebyshev polynomial functions.
%---------------------------------

\begin{figure}[htbp]
	\centering
	\includegraphics[width=1.0\linewidth]{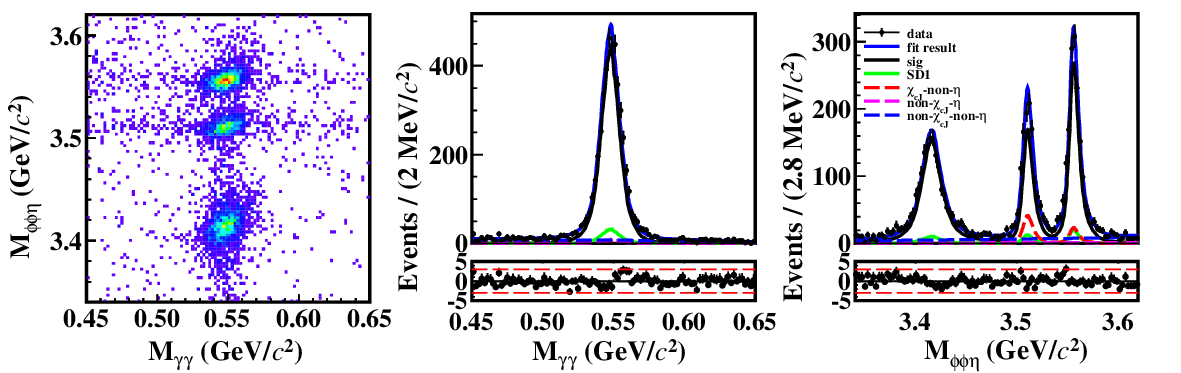}
    \includegraphics[width=1.0\linewidth]{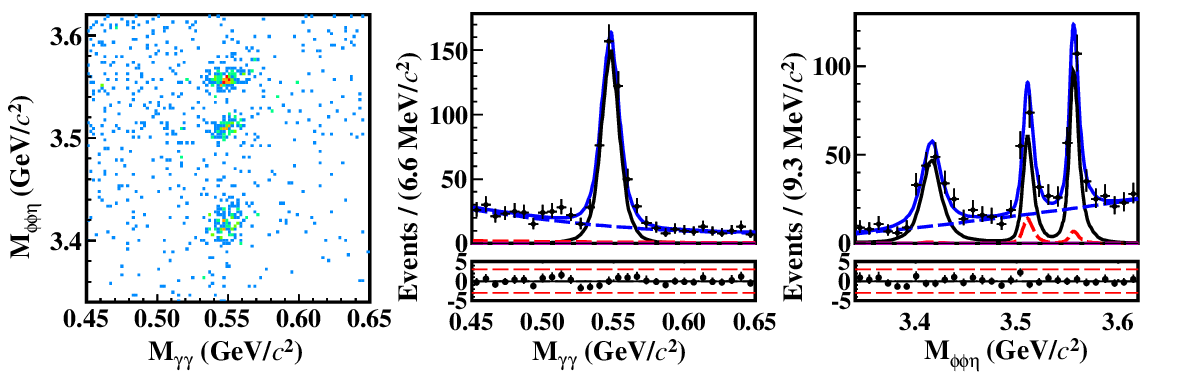}
    \includegraphics[width=1.0\linewidth]{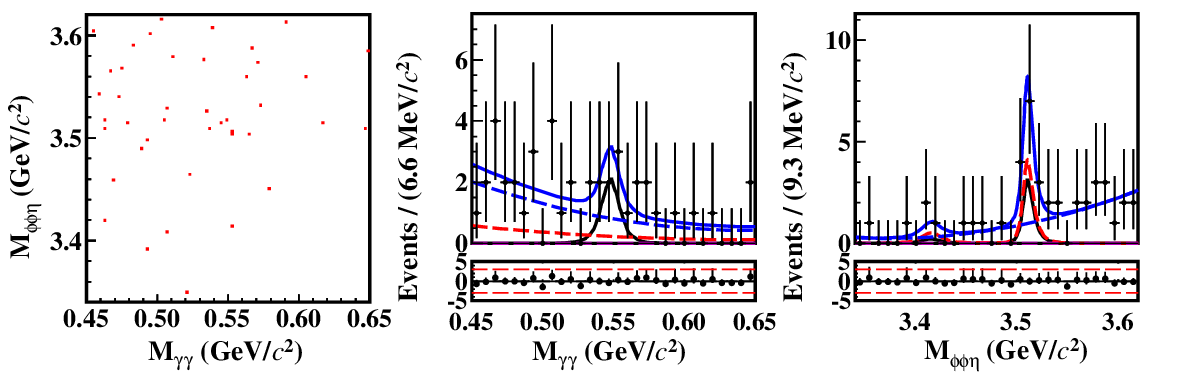}
	\caption{(Left) The distribution of $M_{\phi\phi\eta}$ versus $M_{\gamma\gamma}$ of the accepted $\chi_{cJ} \to \phi\phi\eta$ candidate events in the $\phi\phi$ signal region (first row), SD1 region (second row), and SD2 region (third row).
 (Center and right) The projections of the simultaneous 2D fit on the $M_{\gamma\gamma}$ and $M_{\phi\phi\eta}$ distributions. The dots with error bars are data, the blue solid curves are the fit results, the green solid curves are the normalized sideband contributions, the pink, red, blue dashed curve are the fitted background shapes, and the black solid curves are the fitted signal shapes.}
	\label{fig:fit-1-eta}
\end{figure}

\begin{figure}[htbp]
	\centering
    \includegraphics[width=1.0\linewidth]{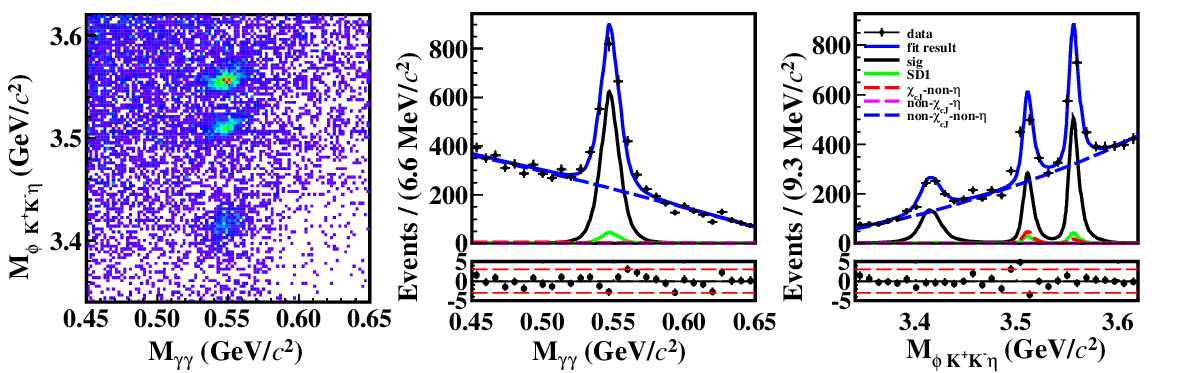}
    \includegraphics[width=1.0\linewidth]{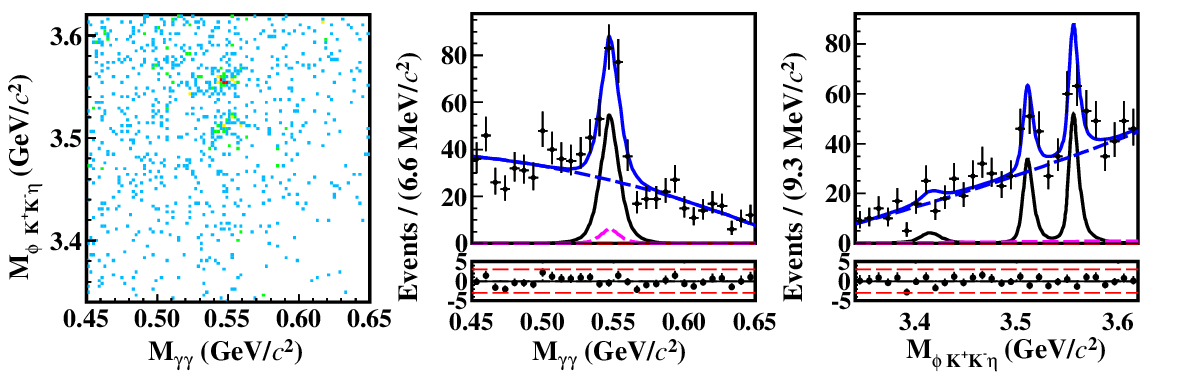}
	\caption{(Left) The distribution of $M_{\phi K^+K^-\eta}$ versus $M_{\gamma\gamma}$ of the accepted $\chi_{cJ} \to \phi K^+K^-\eta$ candidate events in the $\phi$ signal (first row) and $\phi$ sideband (last row) regions.
 (Center and right) The projections of the simultaneous 2D fit on the $M_{\gamma\gamma}$ and $M_{\phi K^+K^-\eta}$ distributions. The dots with error bars are data, the blue solid curves are the fit results, the green solid curves are the normalized sideband contributions, the pink, red, blue dashed curve are the fitted background shapes, and the black solid curves are the fitted signal shapes.}

	\label{fig:fit-2-eta}
\end{figure}

For the decays $\chi_{cJ}\to \phi \phi \eta$ and $\phi K^+K^-\eta$, the fit results are shown in Figs.~\ref{fig:fit-1-eta} and \ref{fig:fit-2-eta}, respectively. The numbers of candidate events for the three $\chi_{cJ}$ decays are summarized in Table~\ref{tab:BF}.

%According to the fitting result, the background contribution from the type-II sideband is negligible, and can be ignored in analysis.

\begin{figure}[htbp]
	\centering
    \includegraphics[width=1.0\linewidth]{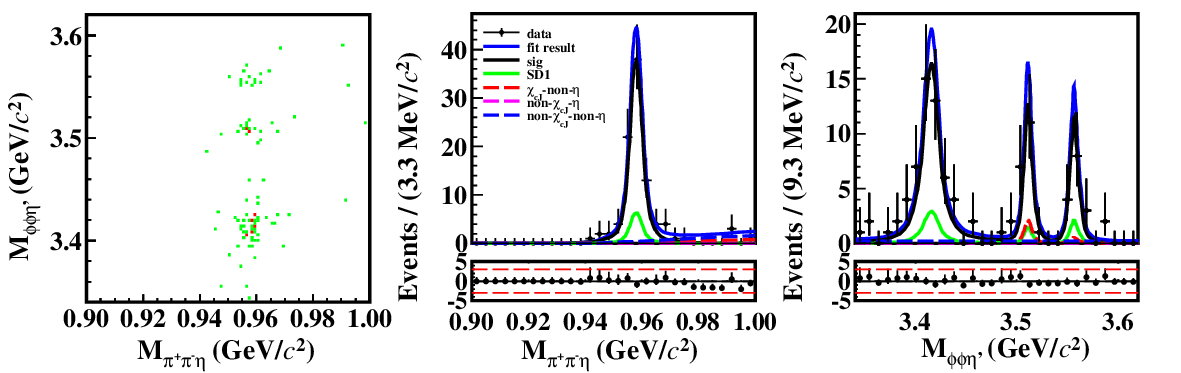}
    \includegraphics[width=1.0\linewidth]{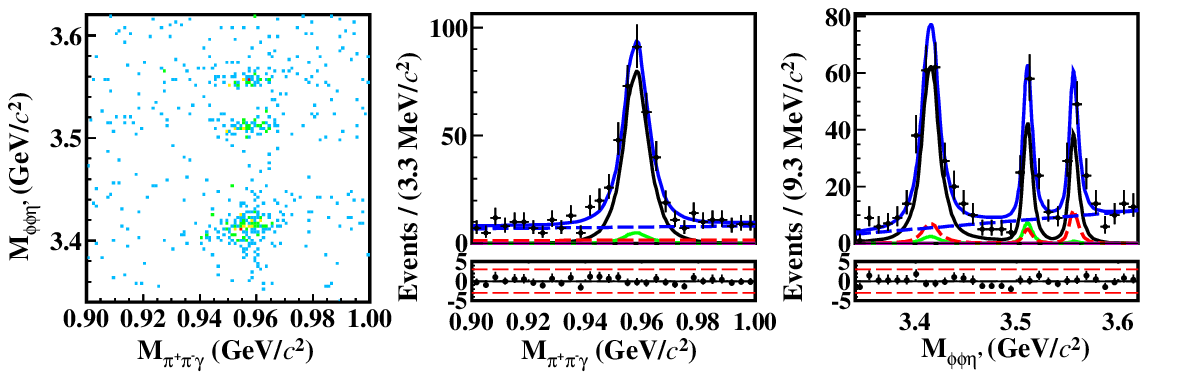}
	\caption{(Left) The distribution of $M_{\phi\phi\eta^{\prime}}$ versus $M_{\pi^{+}\pi^{-}\gamma}$ of the accepted $\chi_{cJ} \to \phi\phi\eta^{\prime}$ candidate events in the $\phi\phi$ signal region with $\eta^{\prime}$ reconstructed in Mode I (first row) and Mode II (last row).
 (Center and right) The projections of the simultaneous 2D fit to the $M_{\pi^{+}\pi^{-}\gamma}$ and $M_{\phi\phi\eta^{\prime}}$ distributions. The dots with error bars are data, the blue solid curves are the fit results, the green solid curves are the normalized sideband contributions, the pink, red, blue dashed curve are the fitted background shapes, and the black solid curves are the fitted signal shapes. The contribution of SD2 is negligible and not shown here.
 }
	\label{fig:fit-1-etap}
\end{figure}

For the decays $\chi_{cJ}\to \phi \phi \eta^{\prime}$, simultaneous 2D unbinned maximum likelihood fits are performed to the distributions of $M_{\phi\phi\eta^{\prime}}$ versus $M_{\eta^{\prime}}$ of the accepted candidate events, with $\eta^{\prime}$ reconstructed via Mode I and Mode II.
%------------------
The fit results are shown in Fig.~\ref{fig:fit-1-etap}, where the results of $\phi\phi$ sideband events are not shown since no significant peaks are found in the fit. After subtracting the sideband contribution, the obtained net yields for all the signal decays are shown in Table~\ref{tab:BF}.
%------------------

The net yields of the signal decays are determined as
%\begin{linenomath*}
\begin{equation}
\begin{aligned}
N_{\rm net} = N_{\mathrm{sig}} - \sum_{i=1, 2} (f_{i}\times N_{\mathrm{SD}, \,i}),
\end{aligned}
\end{equation}	
%\end{linenomath*}
where $N_{\mathrm{sig}}$ and $N_{\mathrm{SD}}$ are the numbers of events in the 2D $\phi\phi$ signal and sideband regions, respectively;
%-------
$f$ is the normalization factor ranging from 0.7 to 1.0, which has been evaluated from the ratios of the background yields in the 2D $\phi\phi$ signal and sideband regions.
%------
Based on these fits, the net signal yields of the $\chi_{c0}$, $\chi_{c1}$ and $\chi_{c2}$ decays are summarized in Table~\ref{tab:BF}.

\begin{figure}[htbp]
  \centering
    \includegraphics[width=0.32\linewidth]{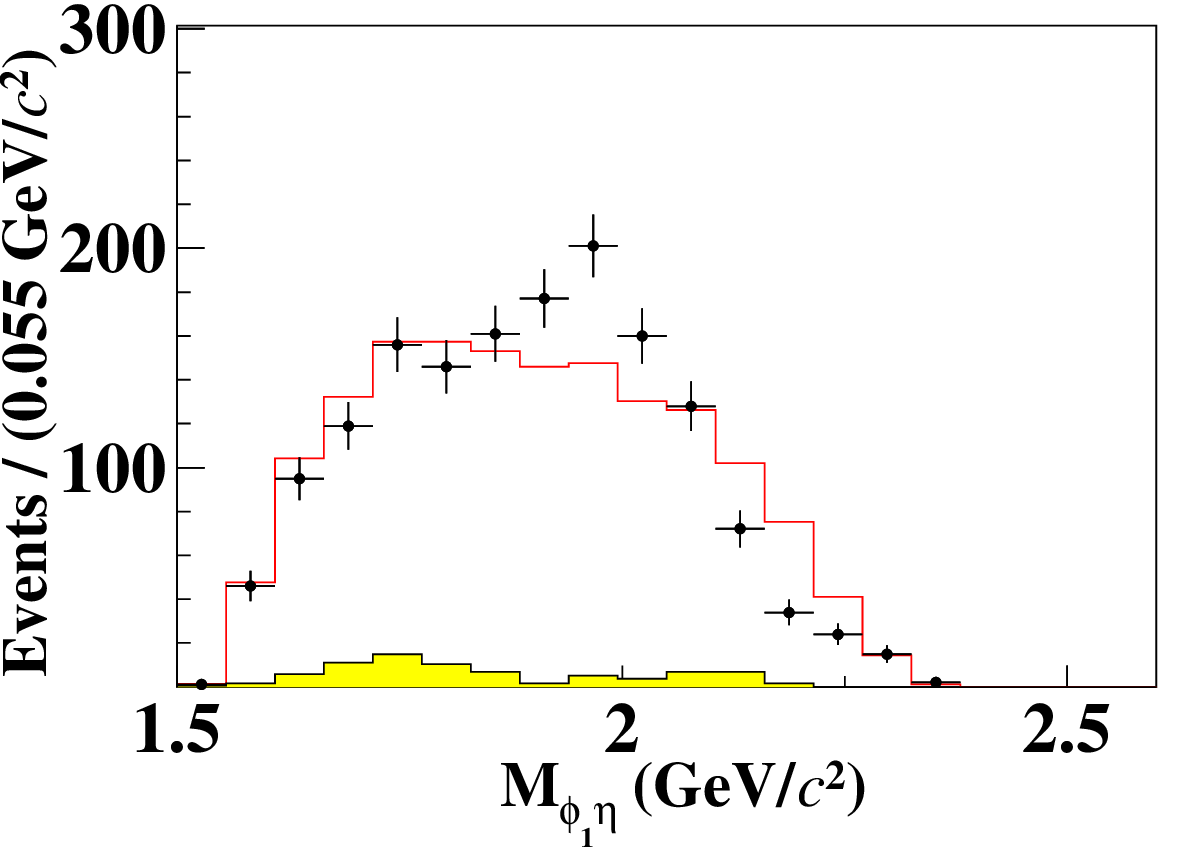}\
    \includegraphics[width=0.32\linewidth]{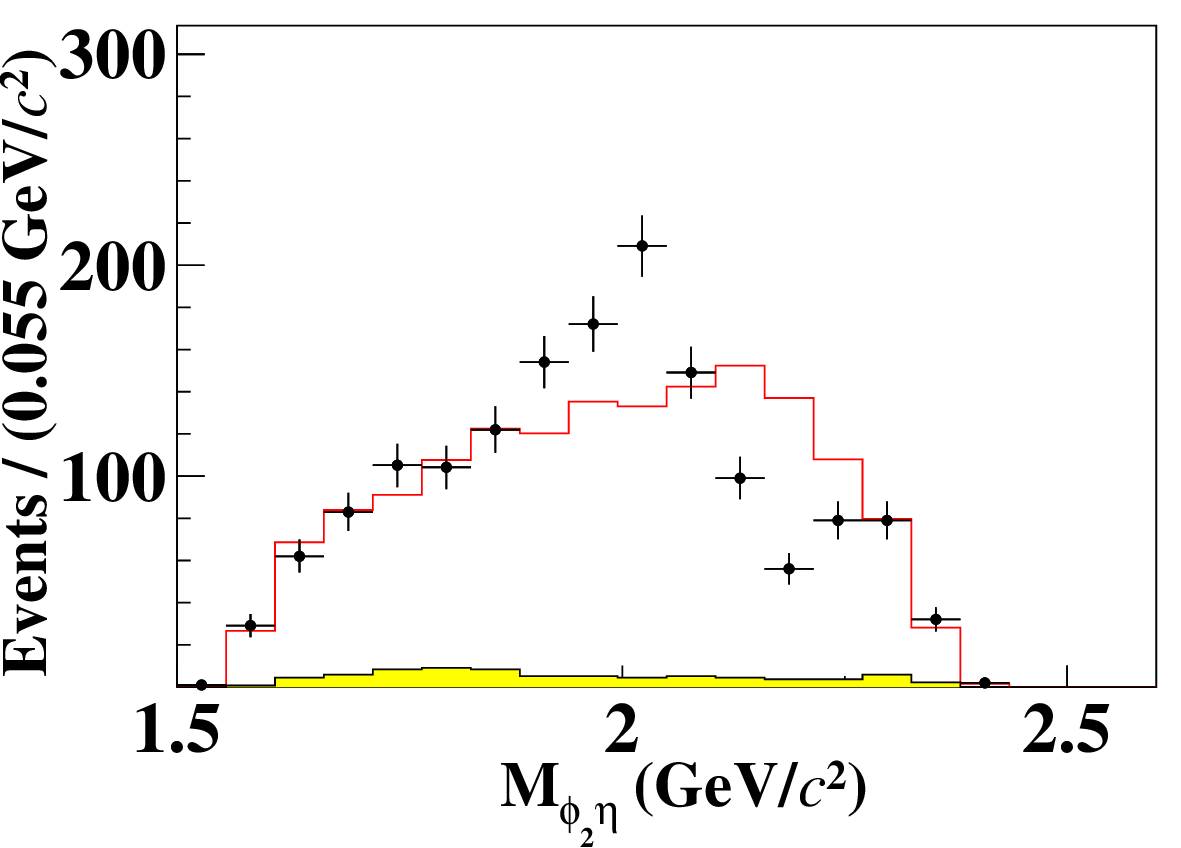}\
    \includegraphics[width=0.32\linewidth]{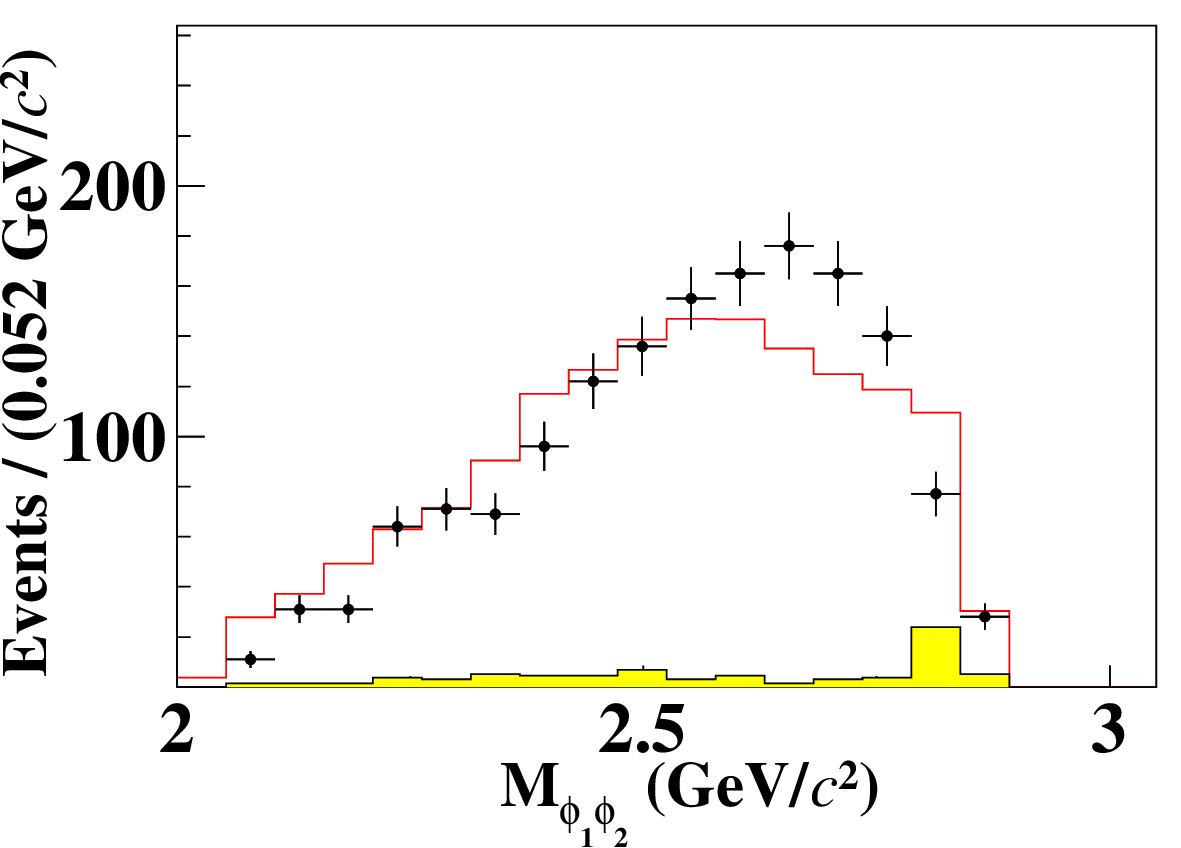}\
    \includegraphics[width=0.32\linewidth]{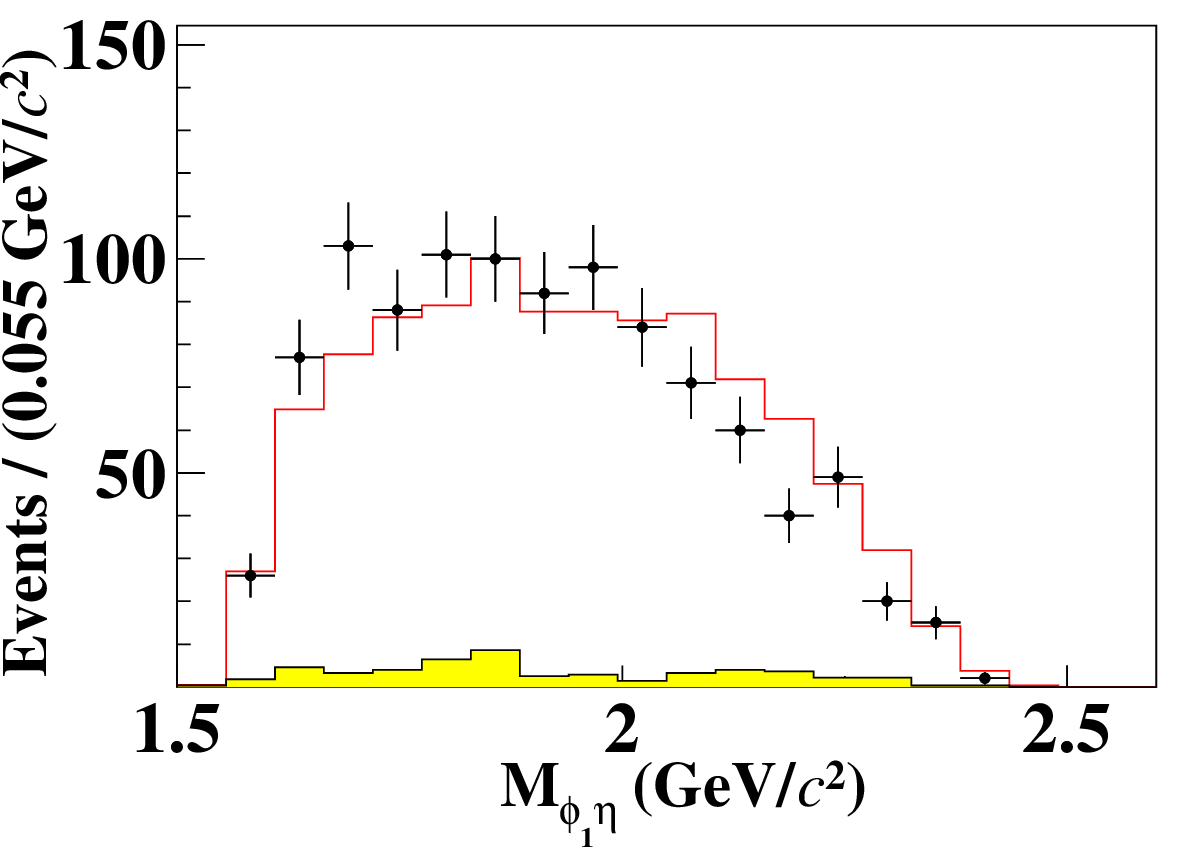}\
    \includegraphics[width=0.32\linewidth]{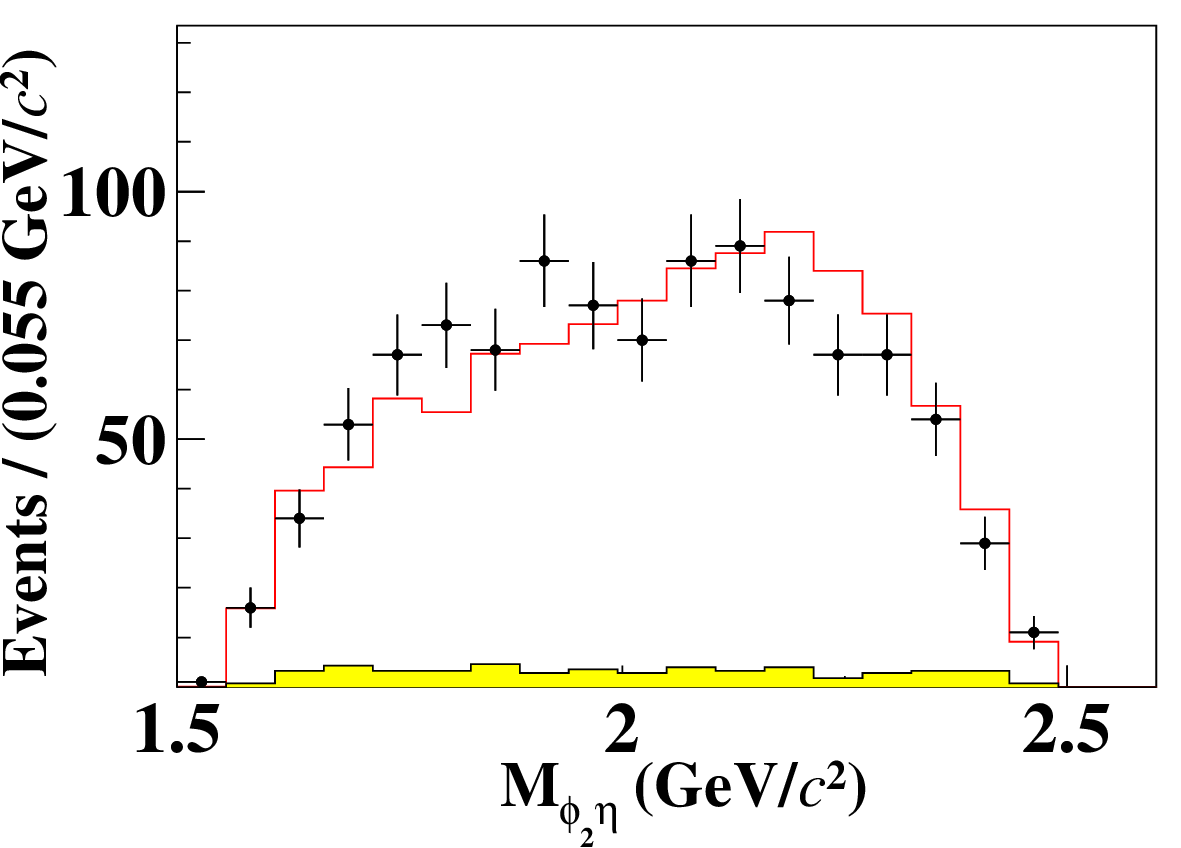}\
    \includegraphics[width=0.32\linewidth]{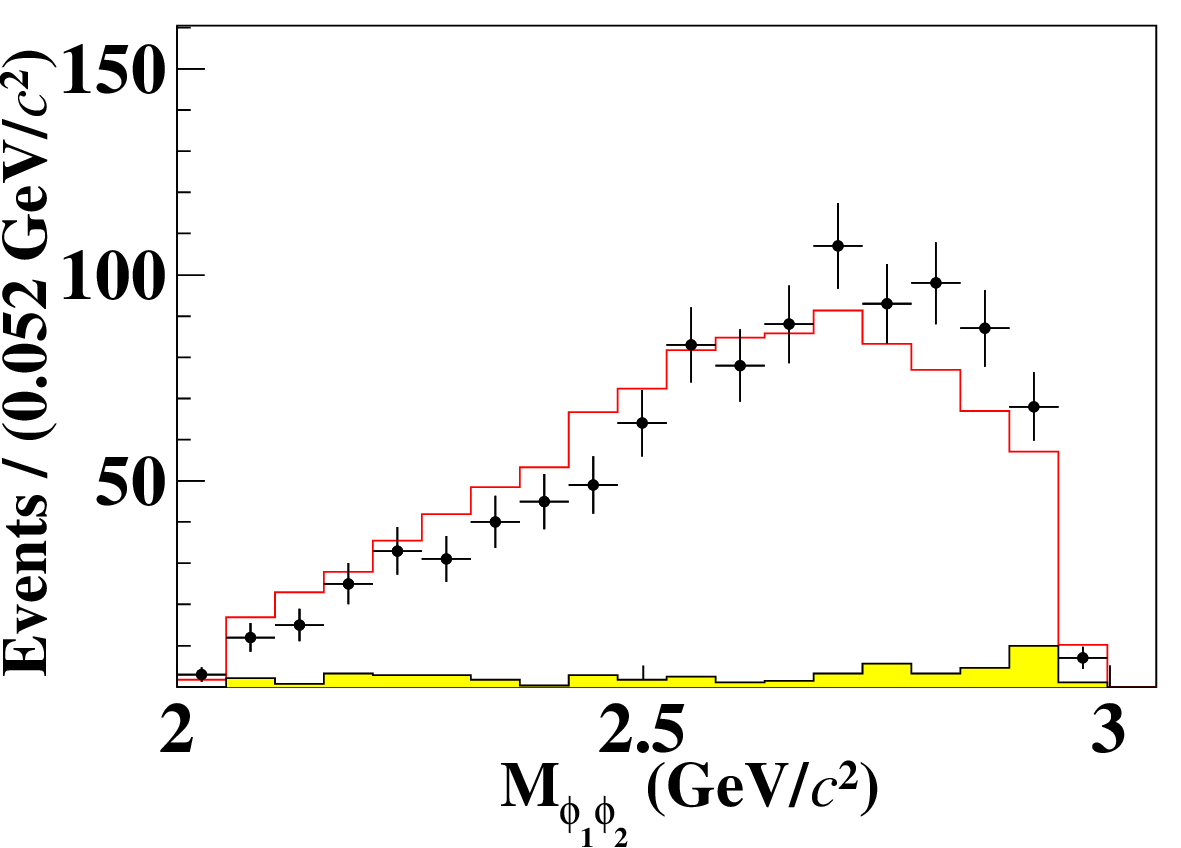}\
    \includegraphics[width=0.32\linewidth]{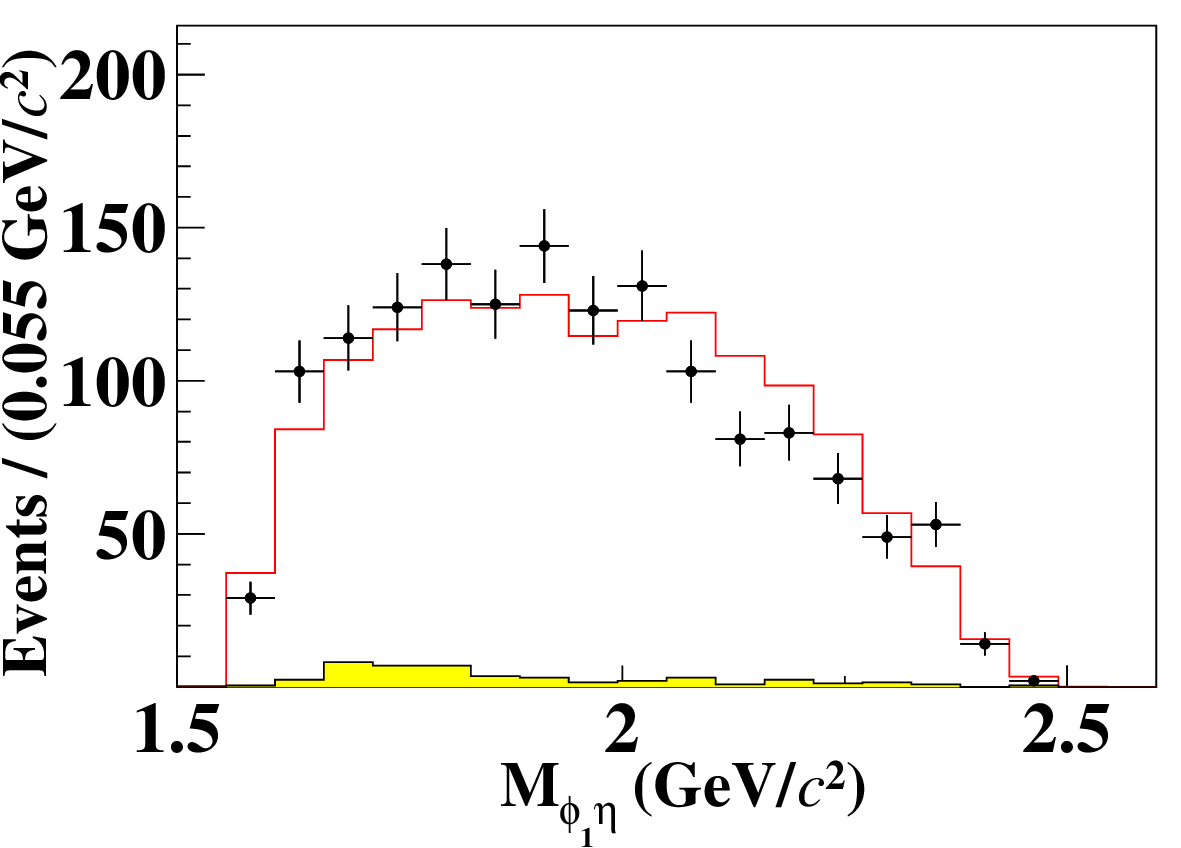}\
    \includegraphics[width=0.32\linewidth]{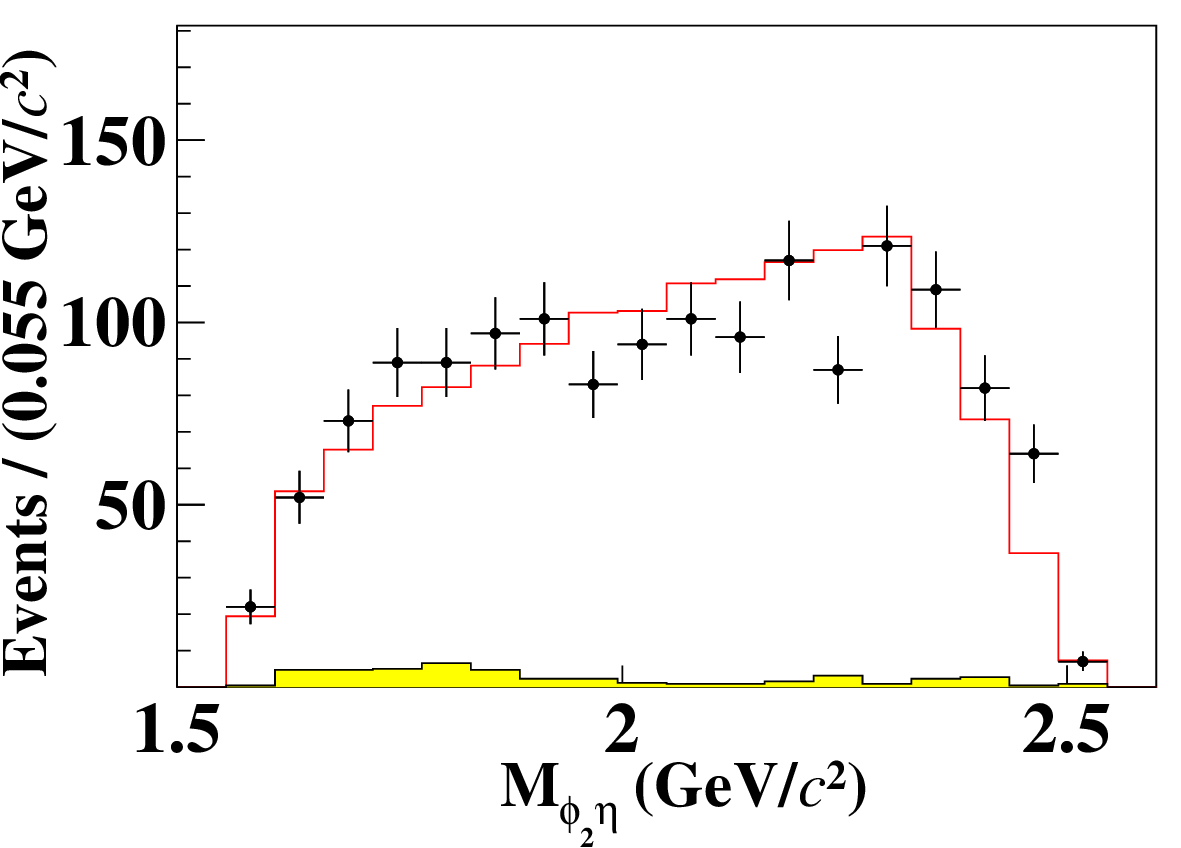}\
    \includegraphics[width=0.32\linewidth]{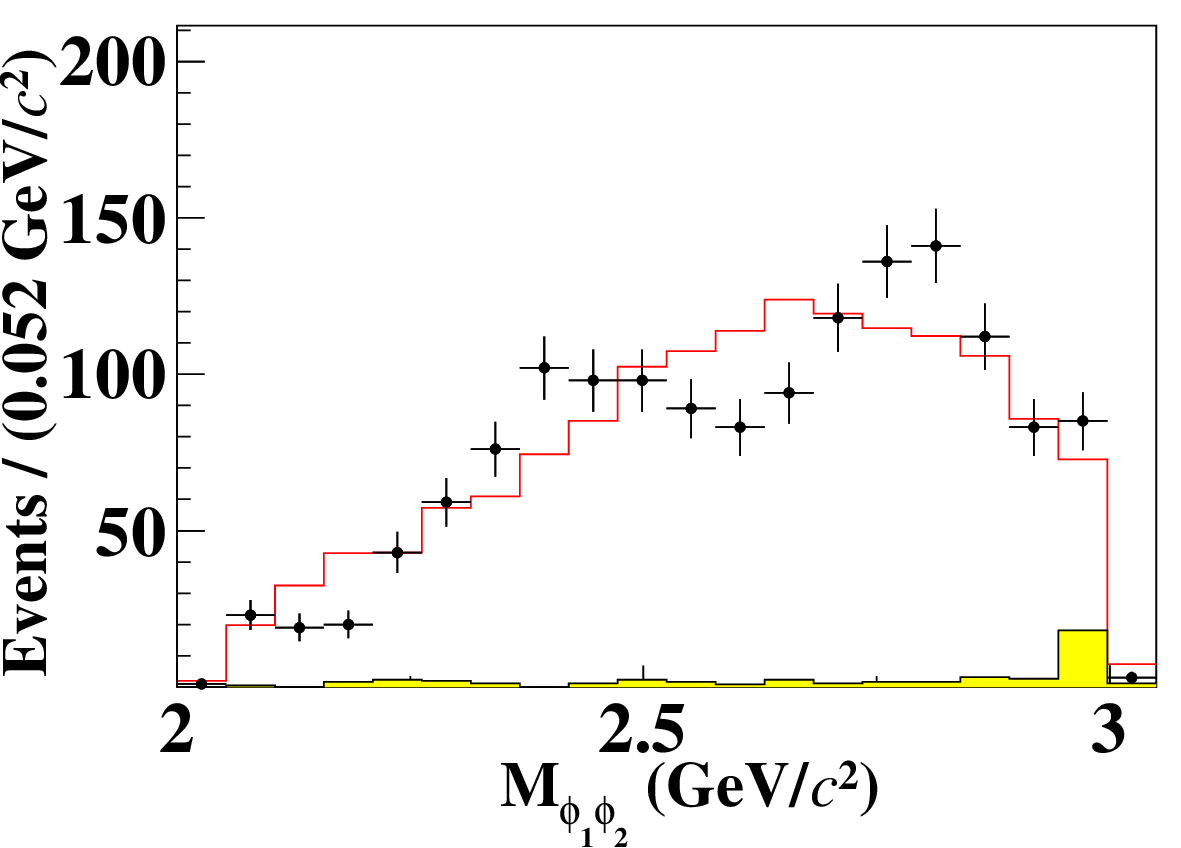}\
  	\caption{Comparisons of the $M_{\phi\phi}$ and $M_{\eta\phi}$ distributions for $\chi_{cJ} \to \phi\phi\eta$ in data (black error bars), signal MC (red histograms), and inclusive MC sample (yellow histograms).
  The top, middle, and bottom rows correspond to $\chi_{c0}$, $\chi_{c1}$, and $\chi_{c2}$, respectively.}
	\label{fig:Compartion-1-eta}
  \end{figure}

\begin{figure}[htbp]
  \centering
    \includegraphics[width=0.32\linewidth]{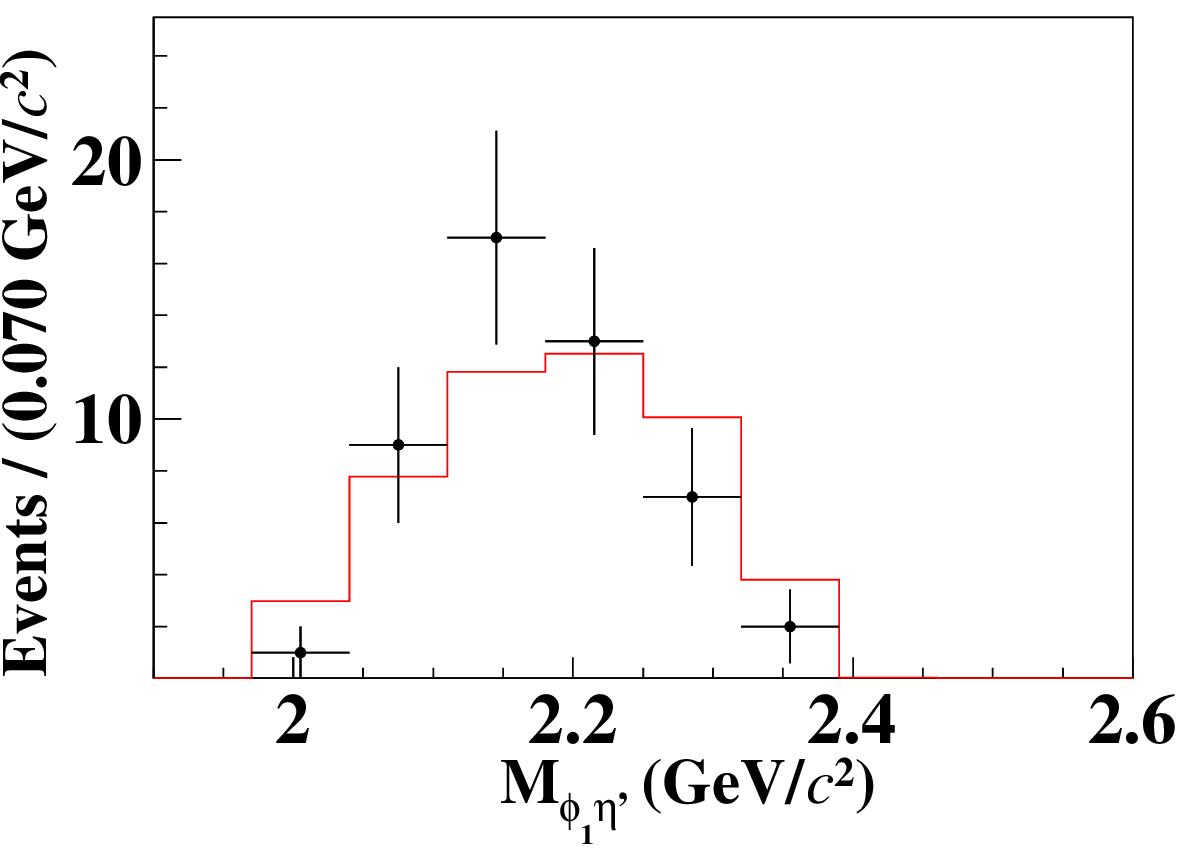}\
    \includegraphics[width=0.32\linewidth]{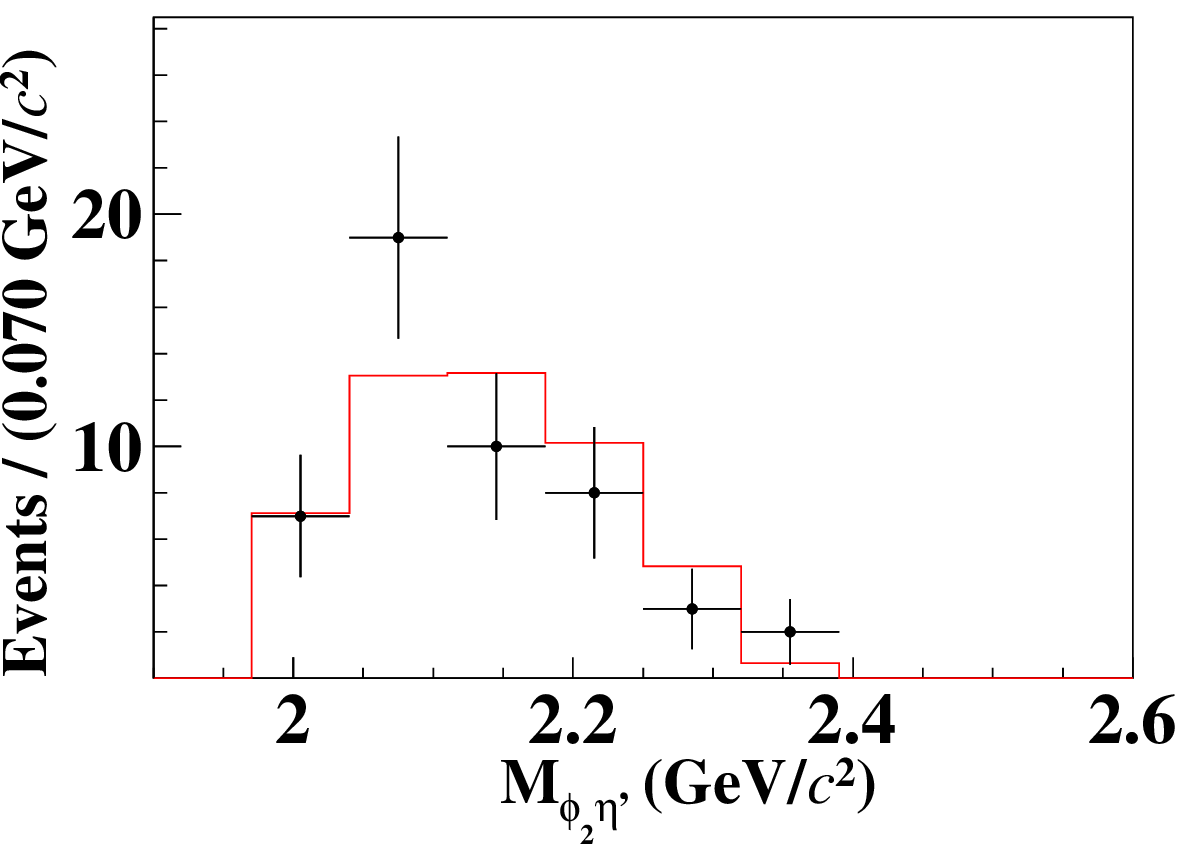}\
    \includegraphics[width=0.32\linewidth]{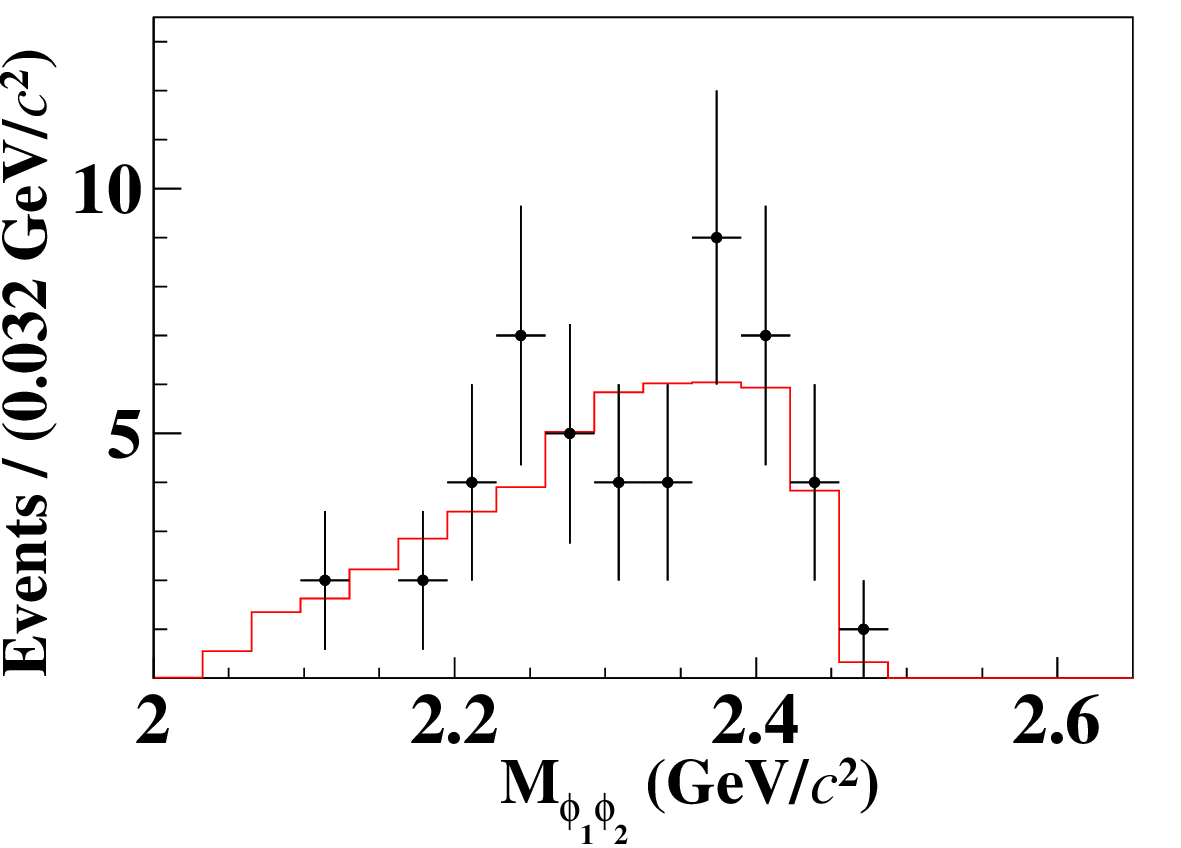}\\
    \includegraphics[width=0.32\linewidth]{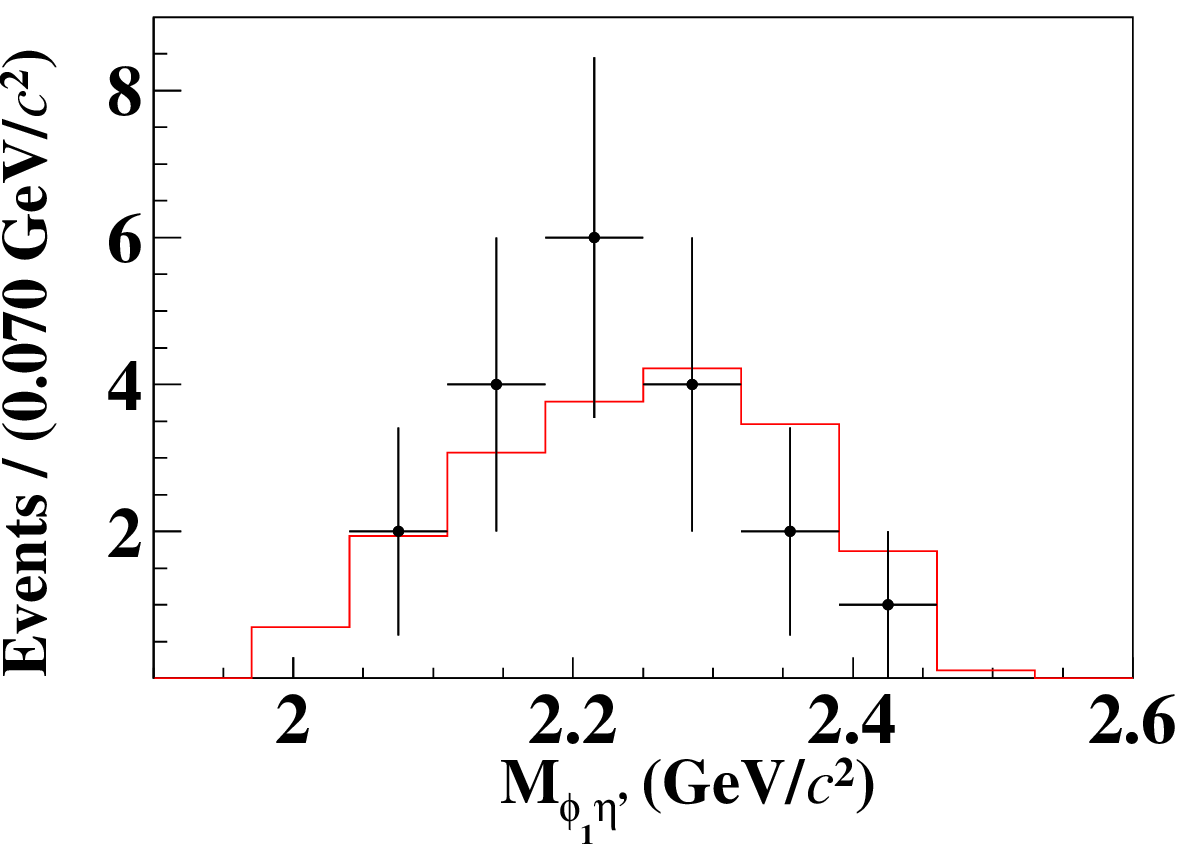}\
    \includegraphics[width=0.32\linewidth]{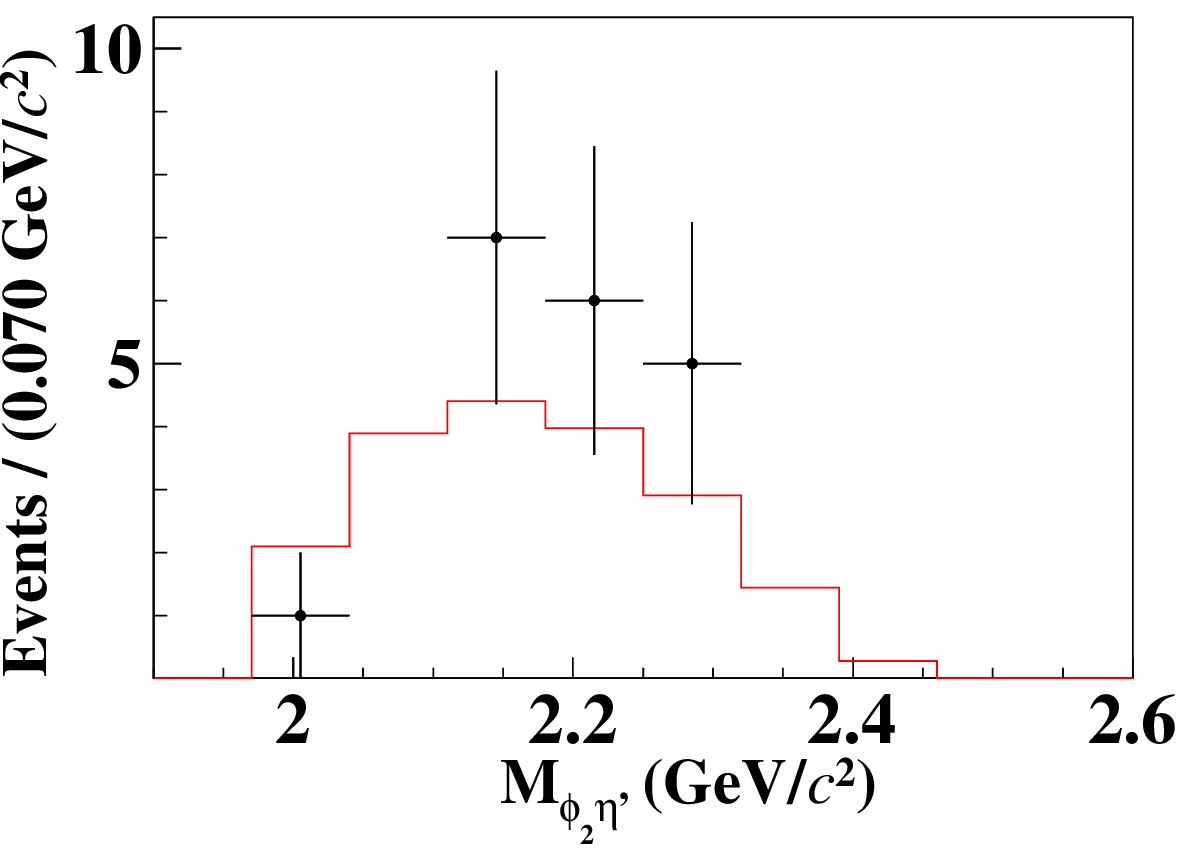}\
    \includegraphics[width=0.32\linewidth]{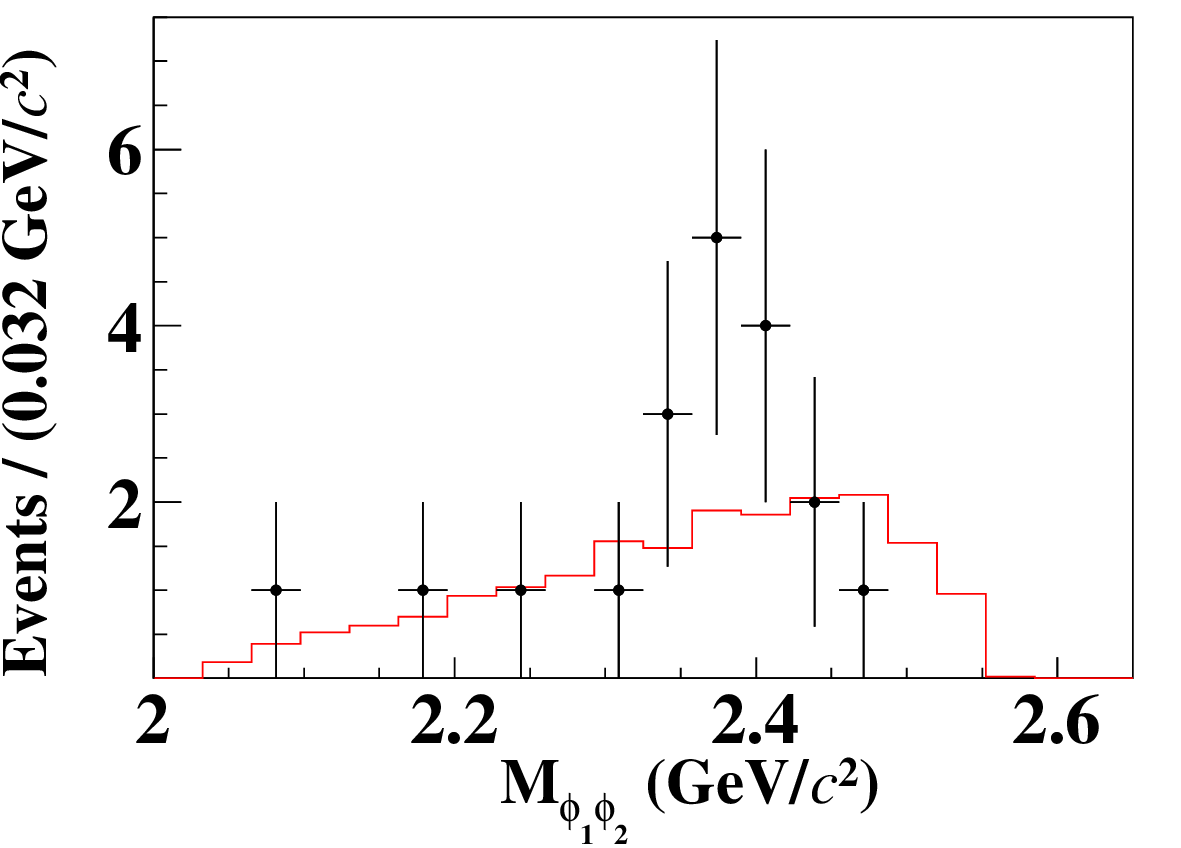}\\
    \includegraphics[width=0.32\linewidth]{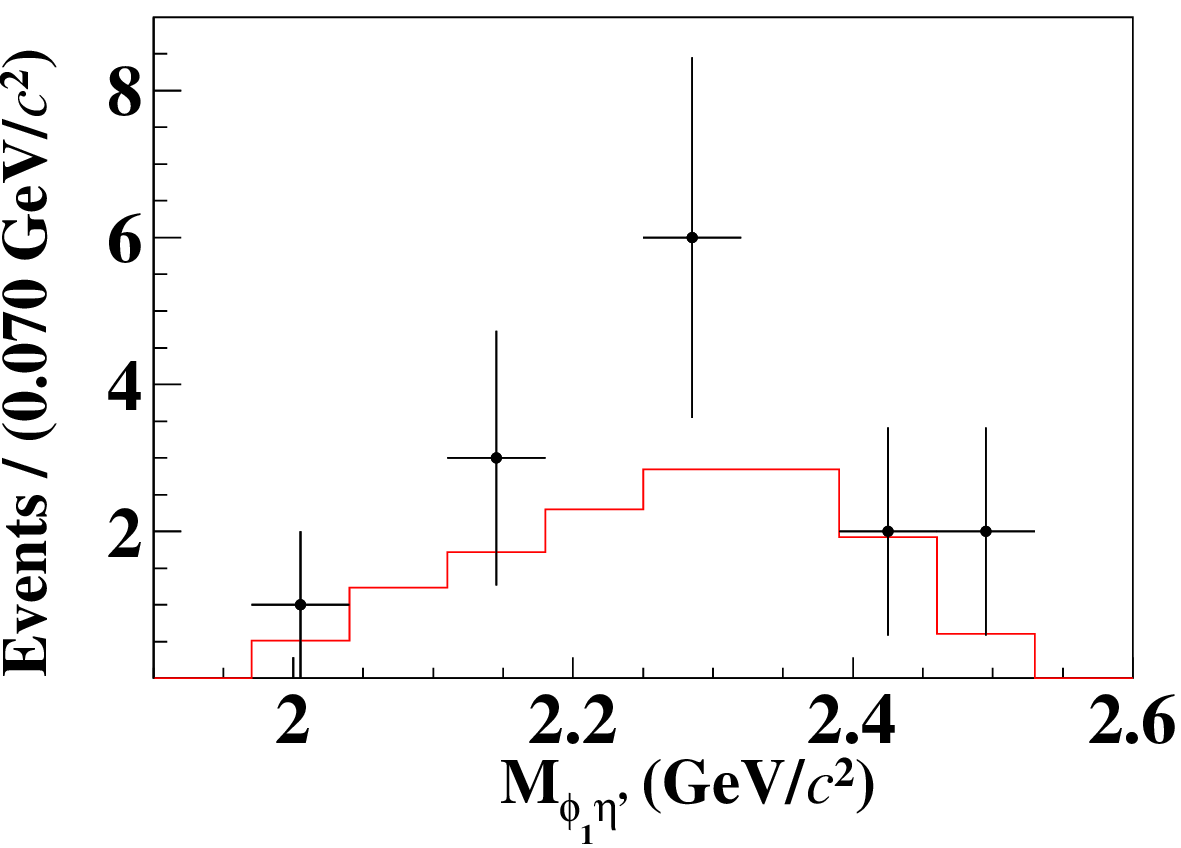}\
    \includegraphics[width=0.32\linewidth]{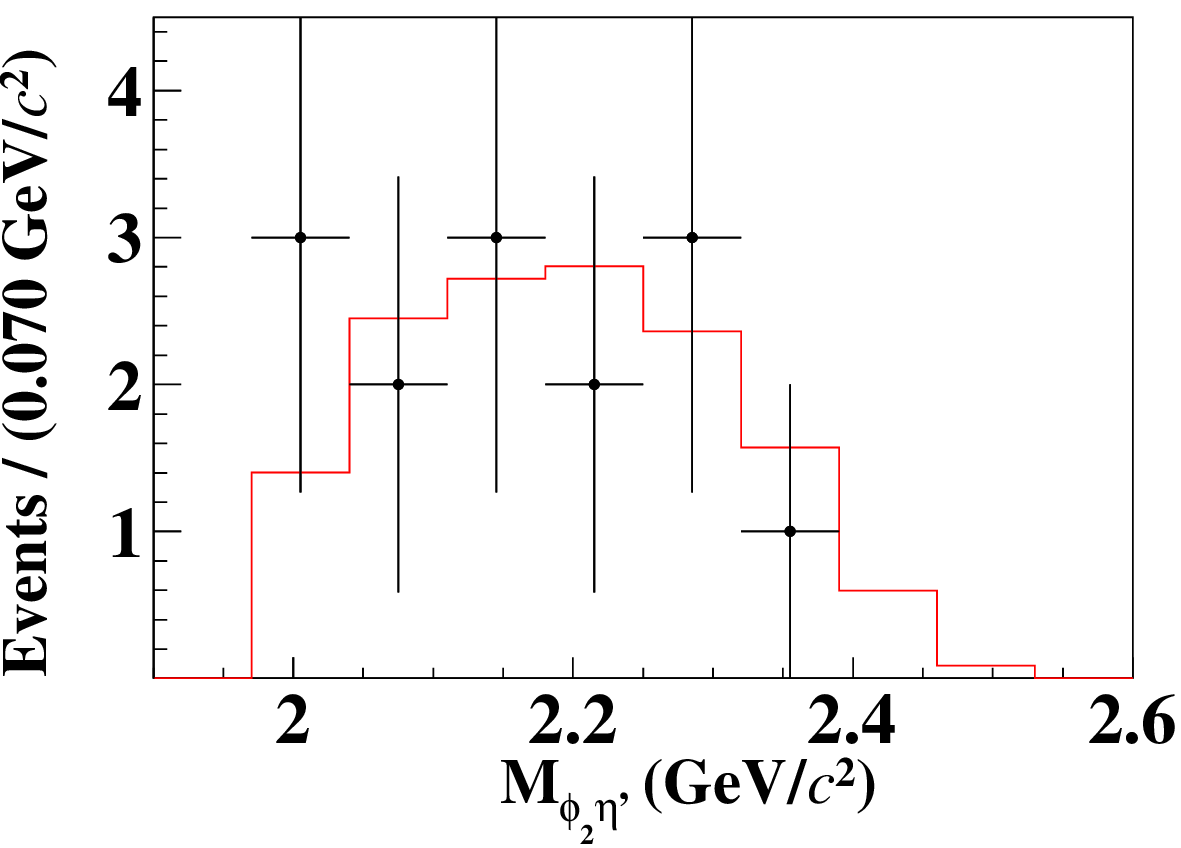}\
    \includegraphics[width=0.32\linewidth]{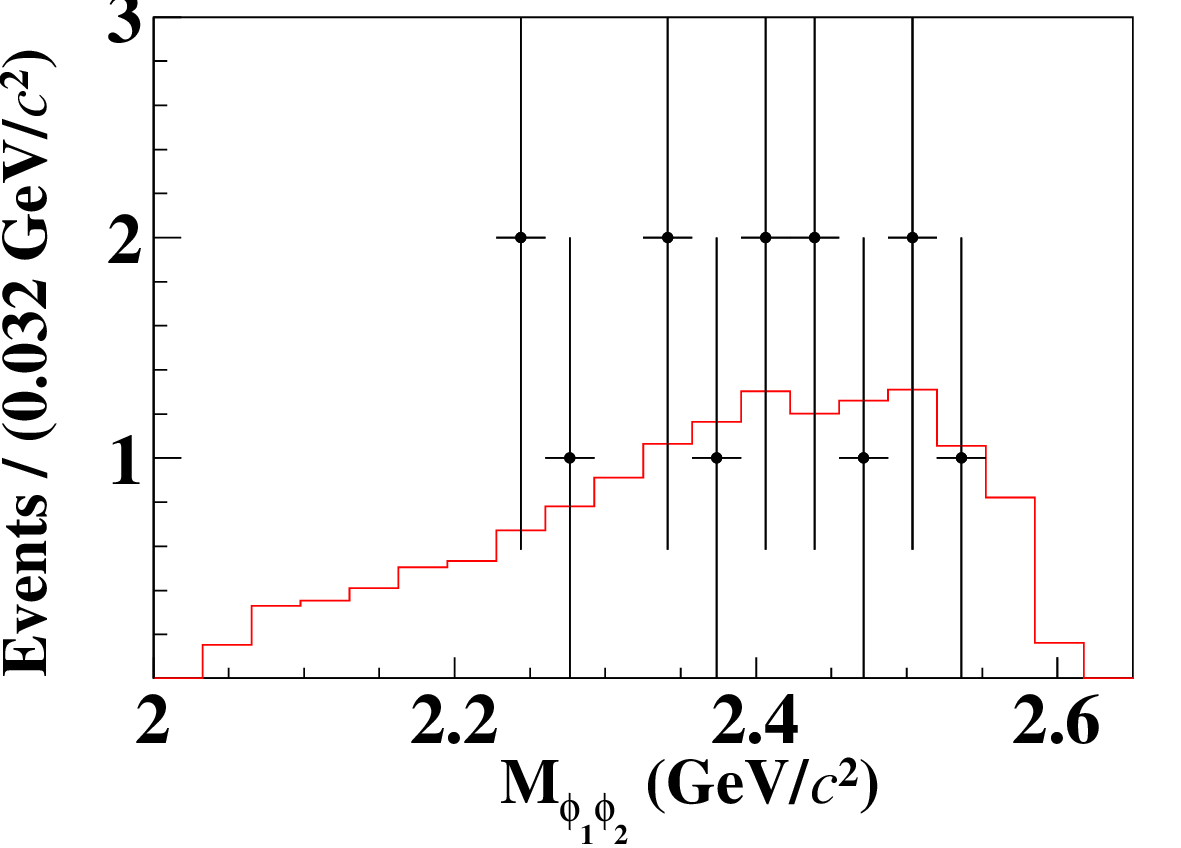}\
  	\caption{Comparisons of the $M_{\phi\phi}$ and $M_{\eta^{\prime}\phi}$ distributions for $\chi_{cJ} \to \phi\phi\eta^{\prime}_{\pi^+\pi^-\eta}$ in data (black error bars) and signal MC sample (red histograms).
   The top, middle, and bottom rows correspond to $\chi_{c0}$, $\chi_{c1}$, and $\chi_{c2}$, respectively.}
	\label{fig:Compartion-1-etap-01}
  \end{figure}

\begin{figure}[htbp]
  \centering
    \includegraphics[width=0.32\linewidth]{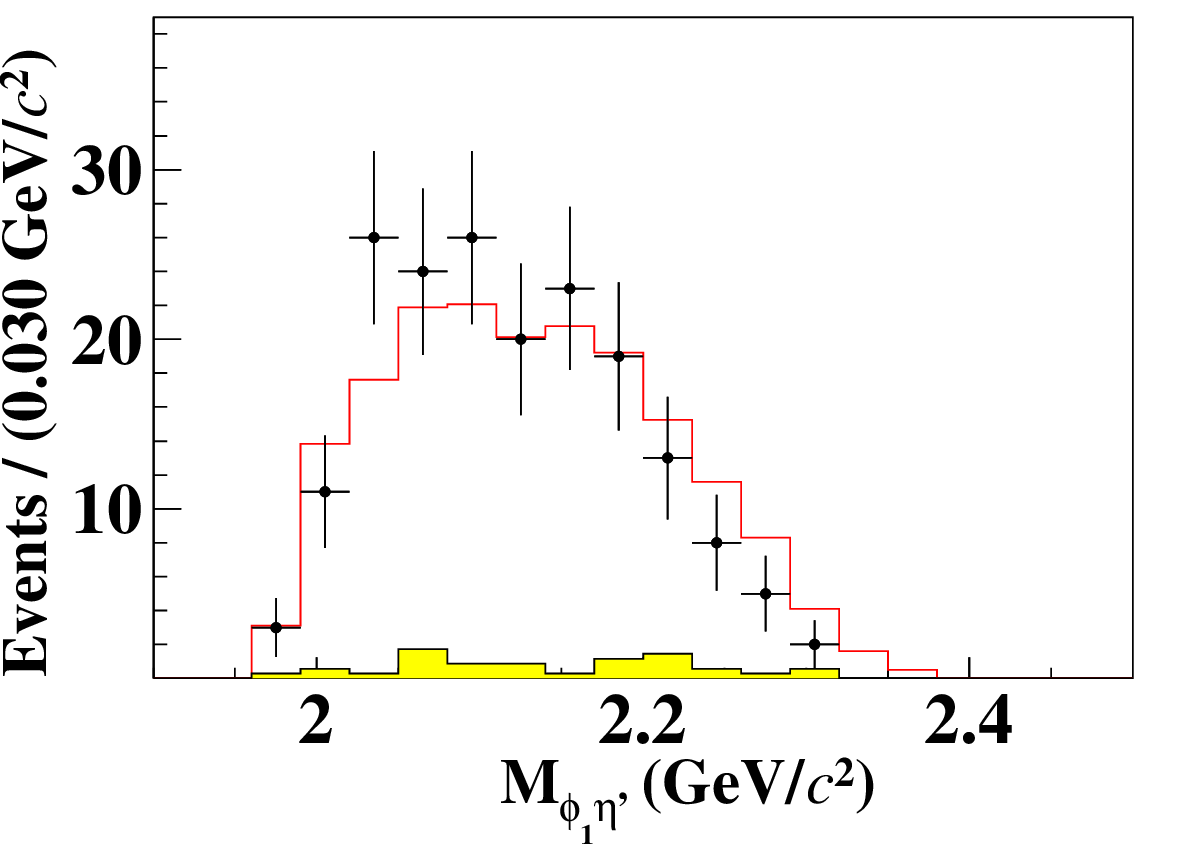}\
    \includegraphics[width=0.32\linewidth]{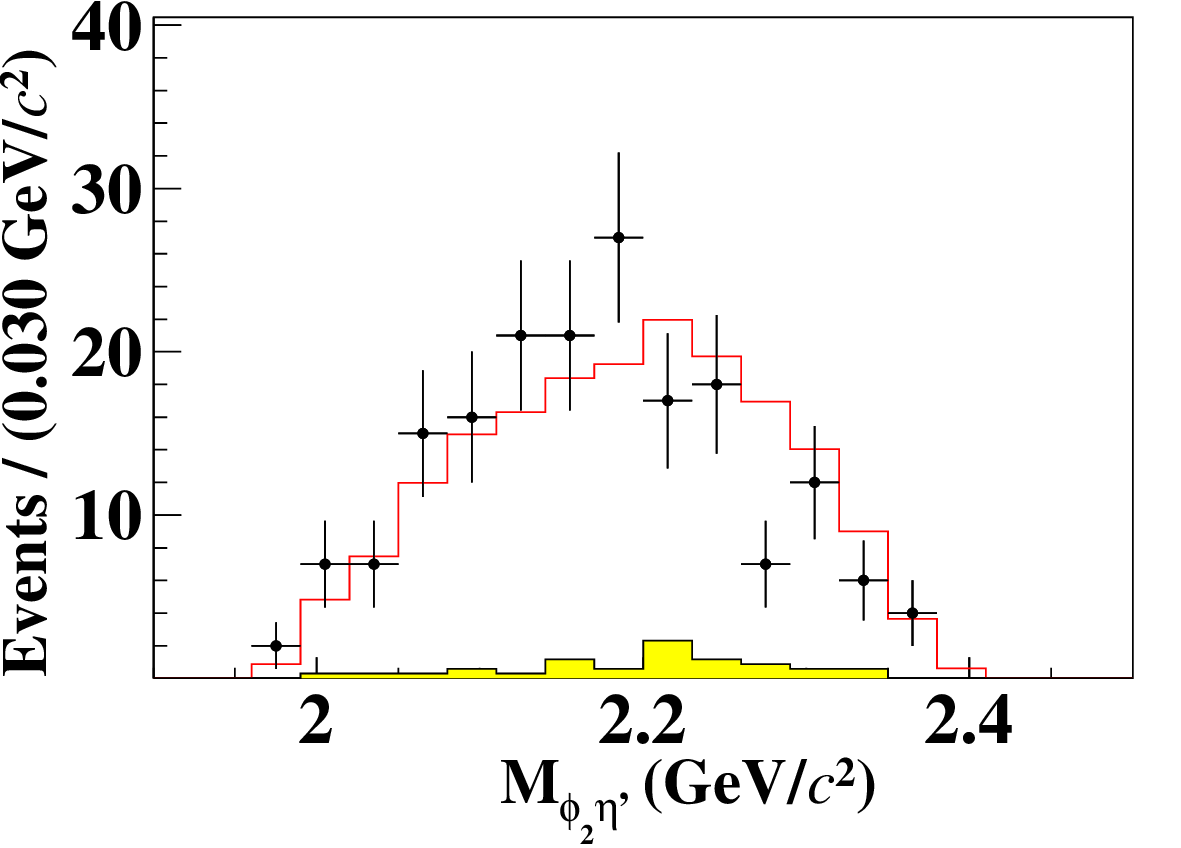}\
    \includegraphics[width=0.32\linewidth]{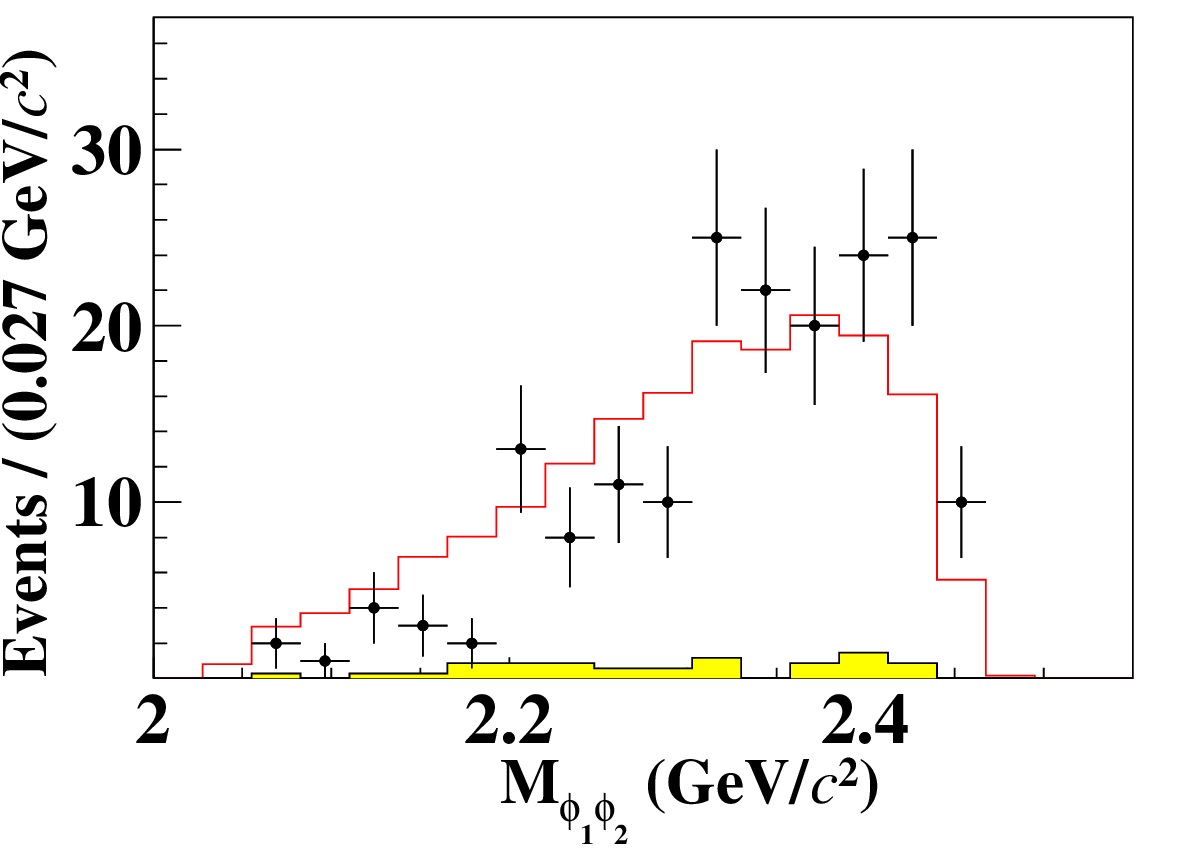}\\
    \includegraphics[width=0.32\linewidth]{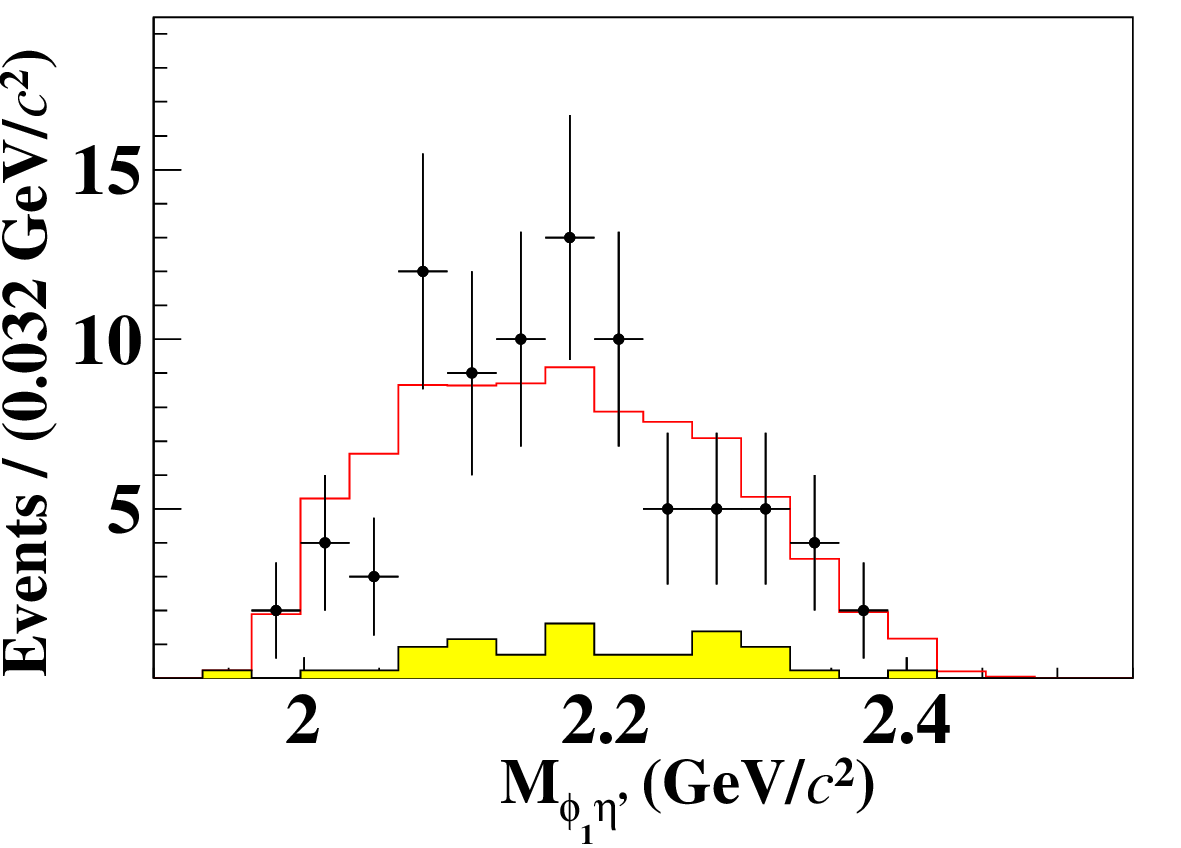}\
    \includegraphics[width=0.32\linewidth]{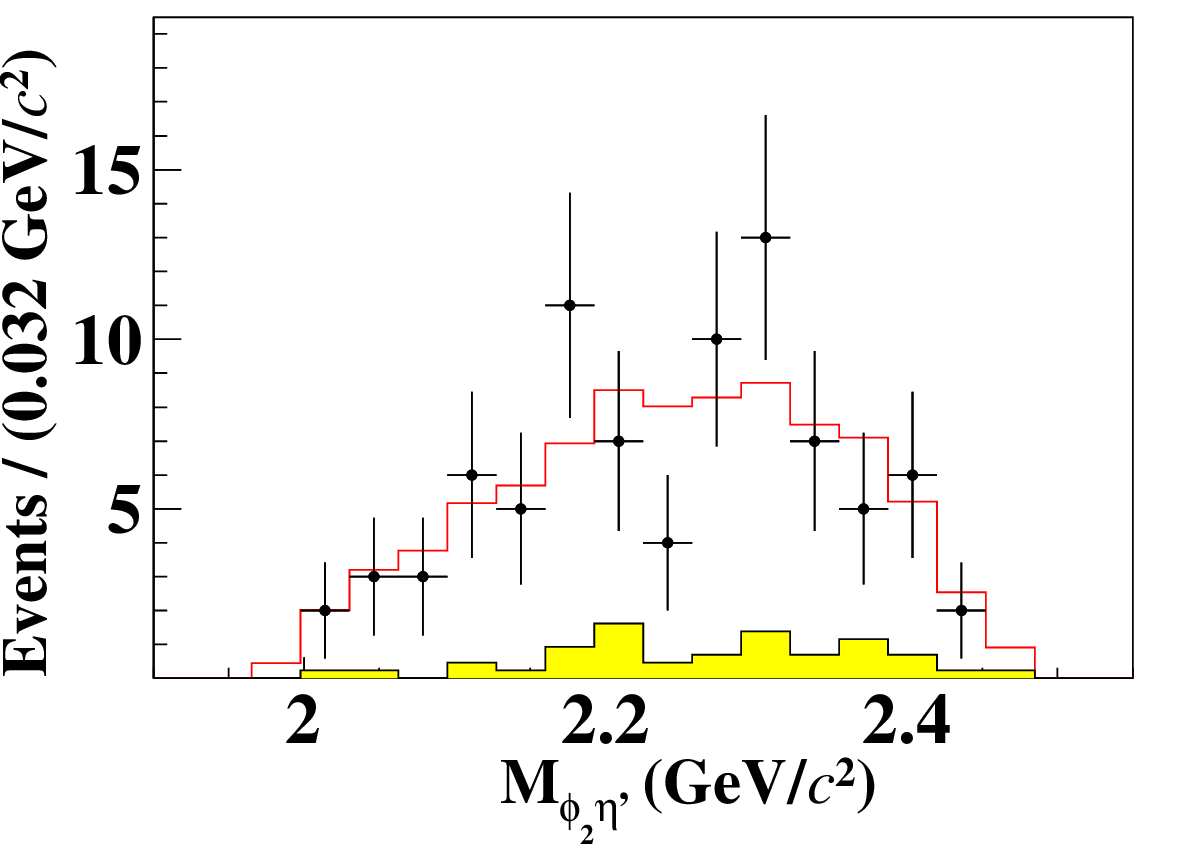}\
    \includegraphics[width=0.32\linewidth]{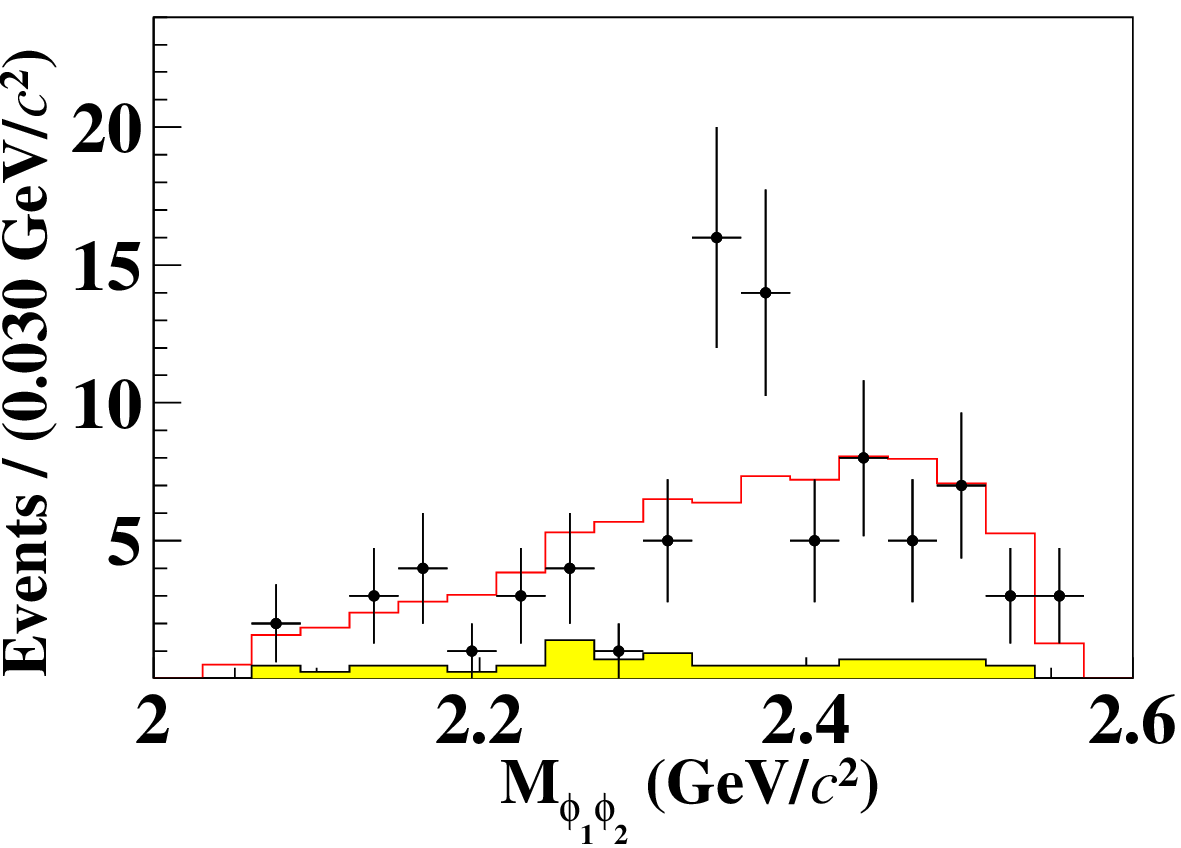}\\
    \includegraphics[width=0.32\linewidth]{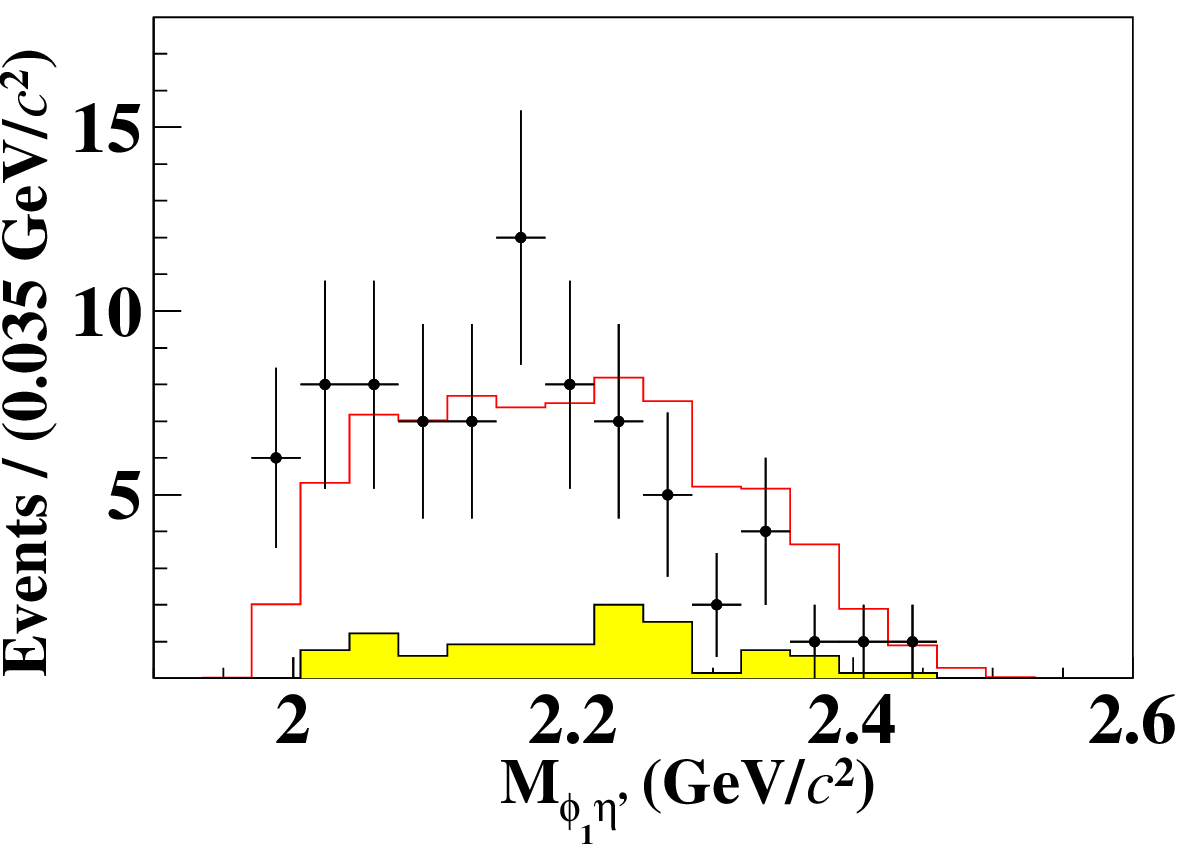}\
    \includegraphics[width=0.32\linewidth]{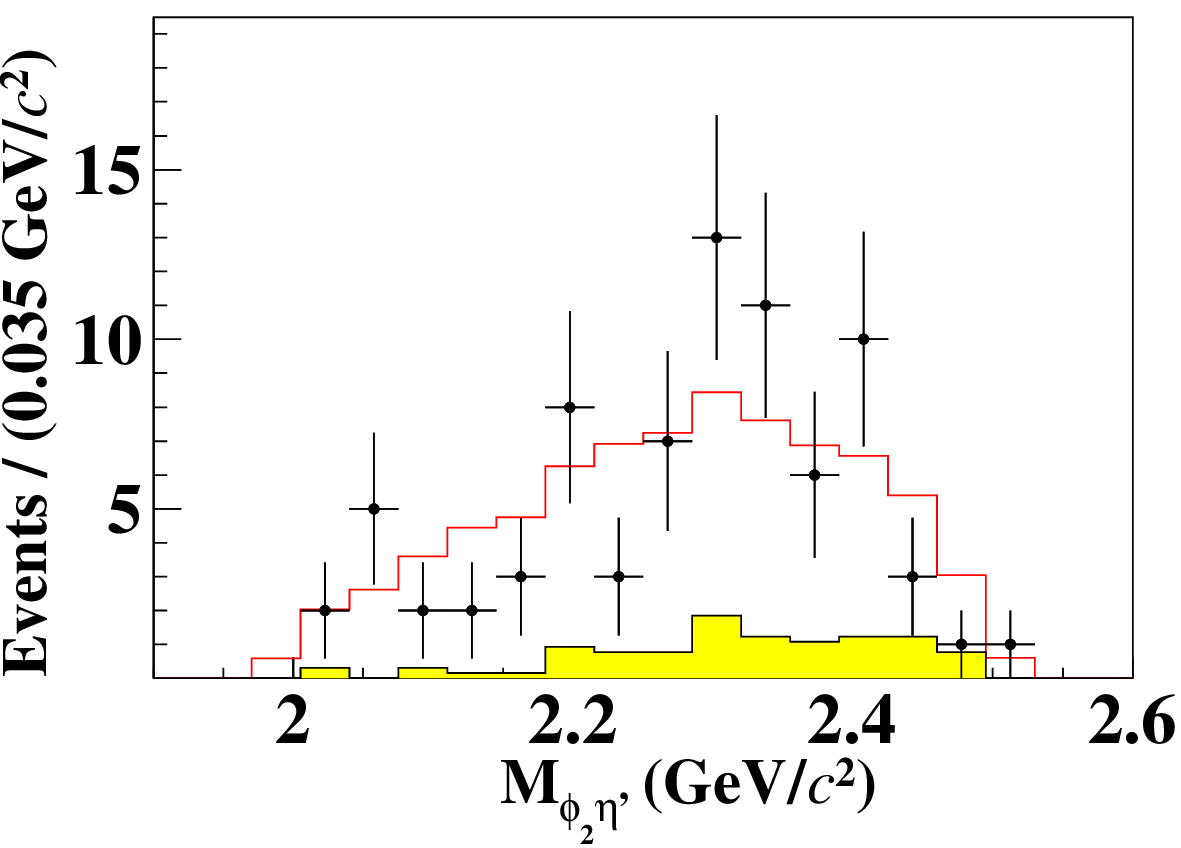}\
    \includegraphics[width=0.32\linewidth]{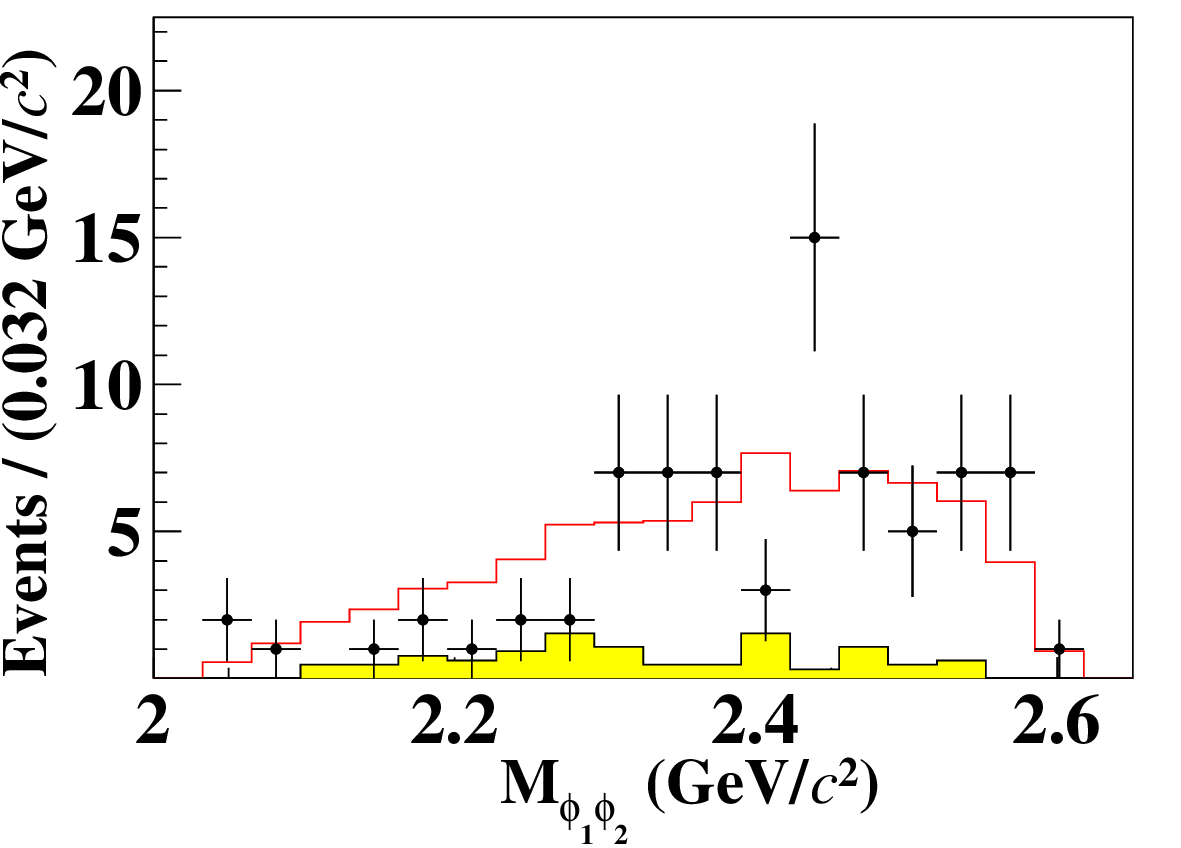}\
  	\caption{Comparisons of the $M_{\phi\phi}$ and $M_{\eta^{\prime}\phi}$ distributions for $\chi_{cJ} \to \phi\phi\eta^{\prime}_{\pi^+\pi^-\gamma}$ in data (black error bars), signal MC (red histograms), and inclusive MC sample (yellow histograms).
   The top, middle, and bottom rows correspond to $\chi_{c0}$, $\chi_{c1}$, and $\chi_{c2}$, respectively.}
	\label{fig:Compartion-1-etap-02}
  \end{figure}

Figures~\ref{fig:Compartion-1-eta},~\ref{fig:Compartion-1-etap-01}, and~\ref{fig:Compartion-1-etap-02} show the comparisons of the distributions of $M_{\phi\phi}$ and $M_{\eta^{(\prime)}\phi}$ of the accepted $\chi_{cJ} \to \phi\phi\eta,~\phi\phi\eta^{\prime}$ candidates in data and signal MC samples.
The data-MC consistency is generally good, thereby ensuring that the signal MC samples are reliable to determine the detection efficiencies.

%\begin{figure}[htbp]
%	\centering
%\includegraphics[width=1.0\linewidth]{figure/4keta/fit/3_4keta.eps}
%\caption{(Left) The distribution of $M_{K^+K^- K^+K^-\eta}$ versus $M_{\eta}$ of the accepted $\chi_{cJ} \to K^+K^+K^-K^-\eta$ candidate events.
% (Middle and right) The projections of the 2D fit on the $M_{K^+K^+K^-K^-\eta}$ and $M_{\eta}$ distributions. The dots with error bars are data, the blue solid curves are the fit results, the pink, red, blue dashed curve are the fitted background shapes, and the black solid curves are the fitted signal shapes.
%}
%	\label{fig:fit-3-eta}
%\end{figure}

 \begin{table*}
 \centering
 \caption{The numbers of events and MC efficiencies for the branching fraction calculations for each signal decay, where the first uncertainties are statistic and the second systematic. The subscriptions 1 and 2 in the last row refer to the Modes I and II used to reconstruct $\eta^{\prime}$, respectively.} %\begin{tabular}{llrrr}
 \begin{tabular}{lcccc}
 \hline\hline % Process  &  & ~~~~~~~~~$\chi_{c0}$~~~~~~~~~ & ~~~~~~~~~$\chi_{c1}$~~~~~~~~~ & ~~~~~~~~~$\chi_{c2}$~~~~~~~~~ \\
    &$\chi_{cJ} \to \phi\phi\eta$ & ~~~~~~~~~~~~$\chi_{c0}$~~~~~~~~~~~~ & ~~~~~~~~~~~~$\chi_{c1}$~~~~~~~~~~~~ & ~~~~~~~~~~~~$\chi_{c2}$~~~~~~~~~~~~ \\
 \hline
% &$N_{\rm sig}$ & $1764\pm50$&$940\pm36$&$1474\pm43$ \\
% &$N_{\rm side1}$ & $152\pm14$&$93\pm11$&$148\pm14$ \\
% &$N_{\rm side2}$ & $1\pm1$&$5\pm3$&$0\pm2$ \\
 &$N_{\rm net}$ & $1652\pm51$&$872\pm37$&$1369\pm44$ \\

 &$\epsilon~(\%)$ & $8.88\pm0.10$&$10.44\pm0.11$&$10.42\pm0.11$ \\
 &$\mathcal B(\psi(3686)\to \gamma\chi_{cJ}) \cdot \mathcal B(\chi_{cJ} \to \phi\phi\eta)~(10^{-5})$  & $7.23\pm0.22\pm0.51$ & $3.25\pm0.14\pm0.23$ & $5.11\pm0.16\pm0.36$ \\
 &$\mathcal B(\chi_{cJ} \to \phi\phi\eta)~(10^{-4})$  & $ 7.40\pm0.23\pm 0.55$ & $3.33\pm0.14\pm0.25$ & $5.46\pm0.17\pm0.40$ \\
 &$\mathcal B(\chi_{cJ} \to \phi\phi\eta)~(10^{-4})$ (PDG)             & $8.4\pm 1.0$  &  $ 3.0\pm0.5$ & $5.4\pm0.7$ \\
 &Significance ($\sigma$)            & $>10$                          & $>10$         &$>10$              \\
 \hline\hline
%-------------------------------------------------------------
    &$\chi_{cJ} \to \phi K^+K^-\eta$ & ~~~~~~~~~~~~$\chi_{c0}$~~~~~~~~~~~~ & ~~~~~~~~~~~~$\chi_{c1}$~~~~~~~~~~~~ & ~~~~~~~~~~~~$\chi_{c2}$~~~~~~~~~~~~ \\
 \hline
%  &$N_{\rm sig}$ & $469\pm29$&$479\pm28$&$842\pm36$ \\
%  &$N_{\rm sideband}$ & $14\pm7$&$52\pm10$&$80\pm11$ \\
  &$N_{\rm net}$ & $456\pm29$&$435\pm30$&$777\pm38$ \\
  &$\epsilon~(\%)$ & $7.23\pm0.09$&$8.53\pm0.10$&$8.68\pm0.10$ \\
  &$\mathcal B(\psi(3686)\to \gamma\chi_{cJ}) \cdot \mathcal B(\chi_{cJ} \to \phi K^+K^-\eta)~(10^{-5})$  & $ 1.20\pm0.08\pm0.09$ & $0.98\pm0.07\pm0.07$ & $1.70\pm0.08\pm0.12$ \\
  &$\mathcal B(\chi_{cJ} \to \phi K^+K^-\eta)~(10^{-4})$  & $ 1.23\pm0.08\pm0.10$ & $1.00\pm0.07\pm0.07$ & $1.82\pm0.09\pm0.14$ \\
  &Significance ($\sigma$)                  &$>10$                             & $>10$           & $>10$             \\
 \hline\hline
%-------------------------------------------------------------
%  %\multirow{3}{*}{$\chi_{cJ} \to K^+K^-K^+K^-\eta$}
%  &$N_{\rm sig}$ & $24\pm9$&$103\pm14$&$129\pm16$ \\
%  &$\epsilon~(\%)$ & $5.59\pm0.08$&$7.05\pm0.09$&$7.22\pm0.09$ \\
%  &$\mathcal B(\psi(3686)\to \gamma\chi_{cJ}) \cdot \mathcal B(\chi_{cJ} \to K^+K^-K^+K^-\eta)~(10^{-6})$  & $ 0.41\pm0.15\pm0.03$ & $1.37\pm0.19\pm0.09$ & $1.68\pm0.21\pm0.16$ \\
%  &$\mathcal B(\chi_{cJ} \to K^+K^-K^+K^-\eta)~(10^{-5})$  & $ 0.42\pm0.15\pm0.03$ & $1.40\pm0.19\pm0.10$ & $1.79\pm0.22\pm0.17$ \\
%  &Significance($\sigma$)             & $3.4$                             & $9.9$                  & $>10$       \\
% \hline\hline
%-------------------------------------------------------------
   &$\chi_{cJ} \to \phi\phi\eta^{\prime}$ & ~~~~~~~~~~~~$\chi_{c0}$~~~~~~~~~~~~ & ~~~~~~~~~~~~$\chi_{c1}$~~~~~~~~~~~~ & ~~~~~~~~~~~~$\chi_{c2}$~~~~~~~~~~~~ \\
   \hline
%  &$N_{\rm net1}$ &  $181\pm14$ &$58\pm8$ &$53\pm8$ \\
%  &$N_{\rm net2}$ &  $51\pm8$  &$17\pm5$ &$16\pm4$ \\
  &$\epsilon_{1}~(\%)$ & $3.25\pm0.04$&$4.44\pm0.04$&$4.48\pm0.04$ \\
  &$\epsilon_{2}~(\%)$ & $1.60\pm0.03$&$2.36\pm0.03$&$2.42\pm0.03$ \\
  &$\mathcal B(\psi(3686)\to \gamma\chi_{cJ}) \cdot \mathcal B(\chi_{cJ} \to \phi\phi\eta^{\prime})~(10^{-5})$  & $ 2.89\pm 0.22\pm0.27$ & $0.67\pm0.10\pm0.07$ &$0.61 \pm0.08\pm0.06 $ \\
  &$\mathcal B(\chi_{cJ} \to \phi\phi\eta^{\prime})~(10^{-4})$  & $ 2.96 \pm 0.23 \pm 0.29$ & $0.69 \pm 0.10 \pm 0.08$ & $0.65 \pm 0.09 \pm 0.07 $ \\
  &Significance ($\sigma$)             & $>10$                             & $9.5$                  & $9.3$       \\
 \hline\hline
  \label{tab:BF}
 \end{tabular}
 \end{table*}

%--------------------------------------------------------------------------------

%--------------------------------------------------------------------------------

%--------------------------------------------------------------------------------

%--------------------------------------------------------------------------------

\section{Branching fraction}

The branching fractions of $\chi_{cJ} \to $ signal ($\phi\phi\eta, \phi\phi \eta^{\prime}$, and $\phi K^+K^-\eta$) are calculated as
\begin{equation}
\begin{aligned}
&\mathcal B_{\psi(3686) \to \gamma\chi_{cJ}} \cdot \mathcal B_{\chi_{cJ} \to \rm{signal}} =\frac{N_{\rm net}}{N_{\psi(3686)} \cdot\epsilon \cdot\mathcal B_{\rm sub}  },
\end{aligned}
\end{equation}	
where $\mathcal{B}_{\psi(3686)\to \gamma\chi_{cJ}}$ is the branching fraction of $\psi(3686) \to \gamma\chi_{cJ}$, $N_{\psi(3686)}$ is the total number of $\psi(3686)$ events in data, $\epsilon$ is the detection efficiency, %-----
$\mathcal B_{\rm sub}$ corresponds to the product of all the branching fractions of the decay chain.
%-------------
All the obtained results are summarized in Table~\ref{tab:BF}.

\section{SYSTEMATIC UNCERTAINTY}
\label{sec:systematics}

The systematic uncertainties in the branching fraction measurements originate from several sources, as summarized in Table~\ref{tab:sys-1-eta}.
For the decays $\chi_{cJ} \to \phi\phi\eta^{\prime}$, the averaged systematic uncertainties for both $\eta^{\prime} \to \pi^+\pi^-\eta$ and $\eta^{\prime} \to \pi^+\pi^-\gamma$ are summarized in Table \ref{tab:sys-1-etap}.
They are estimated and discussed below.

   \begin{table*}[t]
   \centering
   \caption{Relative systematic uncertainties in the measurements of the branching fractions of $\chi_{cJ}\to \phi\phi\eta$ and $\phi K^+K^-\eta$, respectively (in \%).}
    \begin{tabular}{lccc} \hline \hline
       Source            & ~~~~~~~$\chi_{c 0}$~~~~~~~          & ~~~~~~~$\chi_{c 1}$~~~~~~~      & ~~~~~~~$\chi_{c 2}$~~~~~~~                 \\
       \hline $N_{\psi(3686)}$    & 0.5                & 0.5               & 0.5           \\
       Tracking                   & 4.0                & 4.0               & 4.0             \\
       PID                        & 4.0                & 4.0               & 4.0             \\
       Photon reconstruction      & 3.0                & 3.0               & 3.0             \\
       4C kinematic fit           & 0.8/1.7            & 0.7/1.5           & 0.7/1.6           \\
       $\phi$ mass window         & 0.1/0.1            & 0.1/0.1           & 0.1/0.1          \\
       Invariant mass fit         & 1.1/1.3            & 0.4/0.5           & 0.5/0.3           \\
       Fit range                  & 0.8/2.1            & 0.4/0.7           & 0.3/0.1             \\
       MC model                   & 0.5/1.7            & 1.1/0.7           & 1.4/2.4           \\
       MC statistics              & 1.0/1.1            & 0.9/1.0           & 0.9/1.0           \\
       $\mathcal B$($\phi~\to~K^+K^-$)             & 2.0/1.0           & 2.0/1.0      & 2.0/1.0           \\
       $\mathcal B$($\eta~\to~\gamma\gamma$)              & 0.5      & 0.5      & 0.5          \\
       $\mathcal B$($\psi(3686)~\to~\gamma\chi_{cJ}$)        & 2.4      & 2.8      & 2.5           \\
%       $\mathcal B$ from PDG                & 3.2/2.6/2.5        & 3.5/3.0/2.8       & 3.2/2.7/2.5           \\
      \hline Total (without $\mathcal B$($\psi(3686)~\to~\gamma\chi_{cJ}$)) &7.0/7.5  &7.0/6.9  &7.0/7.2\\
      Total (with $\mathcal B$($\psi(3686)~\to~\gamma\chi_{cJ}$))                & 7.4/7.8        & 7.5/7.4       & 7.4/7.6           \\
      \hline \hline
  \end{tabular}
  \label{tab:sys-1-eta}
  \end{table*}

    \begin{table*}[t]
   \centering
   \caption{Relative systematic uncertainties in the measurements of the branching fractions of $\chi_{cJ} \to \phi\phi\eta^{\prime}$ (in \%). The top and bottom parts are correlated and uncorrelated uncertainties, respectively.}
    \begin{tabular}{lccc} \hline \hline
       Source            & ~~~$\chi_{c 0}$~~~          & ~~~$\chi_{c 1}$~~~      & ~~~$\chi_{c 2}$~~~                 \\
       \hline $N_{\psi(3686)}$        &0.5         & 0.5      & 0.5             \\
       Tracking                   & 6.0        & 6.0        & 6.0              \\
       PID                        & 6.0        & 6.0        & 6.0             \\
       Photon reconstruction (from$~\psi(3686)~\to~\gamma\chi_{cJ}$)    & 1.0        & 1.0        & 1.0              \\
       $\eta^{\prime}$ reconstruction     & 1.0         &1.0         &1.0               \\
       Fit range                                  & 0.4      & 2.7      & 1.6             \\
       MC model                     & 1.7      & 1.1      & 2.4           \\
       $\mathcal B$($\phi~\to~K^+K^-$)            & 2.0      & 2.0      & 2.0           \\
       $\mathcal B$($\psi(3686)~\to~\gamma\chi_{cJ}$)        & 2.4      & 2.8      & 2.5           \\   \hline
       %-------------------
%       normalized factor                               & 0.4      & 0.1      & 0.1    \\
%       $N_{\rm sideband}$                              & 1.7      & 4.3      & 2.4         \\
%       Invariant mass fit                              & 0.4      & 0.4      & 0.4        \\
%       Photon reconstruction (from other)              & 0.9      & 0.9      & 0.9              \\
%       $\eta$ reconstruction                           & 0.2      & 0.2      & 0.2         \\
       4C kinematic fit                                & 1.5      & 1.7      & 1.8               \\
       $\phi$ mass window                              & 0.2      & 0.2      & 0.2            \\
       Invariant mass fit                              & 1.8      & 4.3      & 2.4        \\
       MC statistics                                   & 0.8      & 0.7      & 0.7          \\
       $\mathcal B$($\eta^{\prime}~\to~\pi^+\pi^-\eta(\gamma)$)        & 1.1      & 1.1      & 1.1           \\
       $\mathcal B$($\eta~\to~\gamma\gamma$)                      & 0.1      & 0.1      & 0.1           \\
      \hline Total (without $\mathcal B$($\psi(3686)~\to~\gamma\chi_{cJ}$)) &9.4  &10.5  &9.9 \\
       Total (with $\mathcal B$($\psi(3686)~\to~\gamma\chi_{cJ}$))      & 9.7      & 10.9     & 10.2              \\

      \hline \hline
%      \hline $B_{\phi \rightarrow K^+K^-}$                       &      &1.0    &  \\
%      \hline $B_{\eta \rightarrow \gamma\gamma}$                 &      &0.5    &  \\
%      \hline $B_{\psi^\prime \rightarrow \gamma\chi_{cJ}}$       &2.0   &2.5    &2.1  \\
%      \hline $B_{\eta^\prime \rightarrow \eta\pi^+\pi^-}$        &      &1.2    &  \\
  \end{tabular}
  \label{tab:sys-1-etap}
  \end{table*}

The total number of $\psi(3686)$ events in data is measured to be $N_{\psi(3686)}=(2712.4 \pm 14.3)\times 10^6$ using the inclusive
hadronic events, as described in Ref.~\cite{psip-number}. The uncertainty of $N_{\rm \psi(3686)}$ is 0.5\%.

The systematic uncertainties of the tracking or PID for $K^{\pm}$ and $\pi^{\pm}$ are estimated with the control samples of $J/\psi\to K^{*}\bar{K}$ and $J/\psi \to p\bar p \pi^+\pi^-$. They are assigned as 1.0\% for tracking or PID per track~\cite{ref::tracking}.

The systematic uncertainty in the photon detection is assigned as 1.0\% per photon with the control sample of $J/\psi\to\pi^+\pi^-\pi^0$~\cite{ref::gamma-recon},
%----
and the systematic uncertainty of the $\eta$ reconstruction is assigned as $1.0\%$~\cite{eta-recontruction}.
%----------

The systematic uncertainty from the $\eta^{\prime}$ reconstruction comes from the photon detection and $\eta$ reconstruction.
The average systematic uncertainty of these two items are $0.9\%$ and $0.2\%$, respectively.
The total systematic uncertainty is calculated by combining these two effects in quadrature.
%----

The systematic uncertainty from the 4C kinematic fit is estimated by comparing the signal efficiencies after and before the helix parameter correction,
in which the correction factors are taken from the previous study in Ref.~\cite{ref::helixp}.
The changes of the signal efficiencies are assigned as the systematic uncertainties.

The systematic uncertainty of the $\phi$ mass requirement is evaluated with the control sample of each decay channel. We examine the accepted efficiencies of data and MC simulation. The difference of the accepted efficiencies between the data and MC simulation is taken as the systematic uncertainty.

The systematic uncertainty from the fit procedure comes from the background shape and the fit range.
%---------
Different orders of Chebyshev polynomials are used in the alternative fits to describe the background. The largest difference of the signal yield is assigned as the systematic uncertainty.
%-------------
The fit ranges for $\chi_{cJ}$ and $\eta^{(\prime)}$ are varied of 20~MeV to determine the signal yields. The difference of the signal yield is taken as the systematic uncertainty.
%-----------
For each signal decay, the total systematic uncertainty is calculated by combining these two effects in quadrature.

The systematic uncertainties associated with the MC model are evaluated by re-weighting the signal efficiencies to improve data/MC consistency. The efficiencies in each bin of the signal MC are re-weighted based on the real data distribution. The differences between the original and the re-weighted efficiencies are assigned as the uncertainties.
%---------
For the decays $\chi_{c0,1,2} \to \phi\phi\eta$, the systematic uncertainties are $0.5\%, 1.1\%, 1.4\%$ and $1.7\%, 0.7\%, 2.4\%$ for $\chi_{c0,1,2} \to \phi K^+K^-\eta$.

%---------------
The uncertainties due to the limited statistics of the signal MC samples are assigned by
%----
\(\sqrt{\frac{1-\epsilon}{N\epsilon}},\)
%----------
where $\epsilon$ is the detection efficiency and $N$ is the total number of produced signal MC events.

The branching fractions of the $\chi_{cJ}$, $\phi$ and $\eta^{(\prime)}$ decays are quoted from the PDG~\cite{ref::pdg2024}. Their uncertainties are 2.4\%, 2.8\%, 2.5\% for $\psi(3686)\to \gamma\chi_{c0,1,2}$, 1.0\% for $\phi \to K^+K^-$, 1.2\% for $\eta^{\prime} \to \pi^+\pi^-\eta$, 1.4\% for $\eta^{\prime} \to \pi^+\pi^-\gamma$, and 0.5\% for $\eta \to \gamma\gamma$.

For each signal decay, the total systematic uncertainty is obtained by adding all systematic uncertainties in quadrature.

\section{Summary}

 Utilizing $(2712.4 \pm 14.3)\times 10^6$ $\psi(3686)$ events collected with the BESIII detector,
 the branching fractions of $\chi_{cJ}\to \phi\phi\eta$ are measured with precision improved by factors of $1.5-1.9$ compared to the previous measurement~\cite{BESIII:2019ngm}. The decays $\chi_{cJ}\to \phi\phi \eta^{\prime}$ and $\phi K^+K^-\eta $ are measured for the first time.
 These measurements of the decays $\chi_{cJ} \to VVP$ provide valuable insights into the $\chi_{cJ}$ decay mechanisms. Based on the events in the signal regions, no obvious resonant structure is observed in the $M_{\phi\phi}$ or $M_{\eta^{(\prime)}\phi}$ distribution. Further studies with high statistics data collected at the future super tau-charm factory~\cite{ref::supert} will be crucial for a deeper understanding of the $\chi_{cJ}$ decay and to seek potential new structures in the $M_{\phi\phi}$ or $M_{\eta^{(\prime)}\phi}$ distribution.

\section{Acknowledgement}
%% Saved at => 2025-07-28

The BESIII Collaboration thanks the staff of BEPCII (https://cstr.cn/31109.02.BEPC) and the IHEP computing center for their strong support. This work is supported in part by National Key R\&D Program of China under Contracts Nos. 2023YFA1606000, 2023YFA1606704; National Natural Science Foundation of China (NSFC) under Contracts Nos. 11635010, 11935015, 11935016, 11935018, 12025502, 12035009, 12035013, 12061131003, 12192260, 12192261, 12192262, 12192263, 12192264, 12192265, 12221005, 12225509, 12235017, 12361141819; the Chinese Academy of Sciences (CAS) Large-Scale Scientific Facility Program; the Strategic Priority Research Program of Chinese Academy of Sciences under Contract No. XDA0480600; CAS under Contract No. YSBR-101; 100 Talents Program of CAS; The Institute of Nuclear and Particle Physics (INPAC) and Shanghai Key Laboratory for Particle Physics and Cosmology; ERC under Contract No. 758462; German Research Foundation DFG under Contract No. FOR5327; Istituto Nazionale di Fisica Nucleare, Italy; Knut and Alice Wallenberg Foundation under Contracts Nos. 2021.0174, 2021.0299; Ministry of Development of Turkey under Contract No. DPT2006K-120470; National Research Foundation of Korea under Contract No. NRF-2022R1A2C1092335; National Science and Technology fund of Mongolia; Polish National Science Centre under Contract No. 2024/53/B/ST2/00975; STFC (United Kingdom); Swedish Research Council under Contract No. 2019.04595; U. S. Department of Energy under Contract No. DE-FG02-05ER41374

%% ends here %%

\end{document}